\documentclass[twocolumn,showpacs,preprintnumbers,amsmath,amssymb,prb,superscriptaddress]{revtex4-1}
\usepackage{amsfonts}
\usepackage[english]{babel}
\usepackage[T1]{fontenc}
\usepackage{times}
\usepackage{mathrsfs}
\usepackage{graphicx}
\usepackage{dcolumn}
\usepackage{bm}
\usepackage[colorlinks,bookmarks=true,citecolor=blue,linkcolor=red,urlcolor=blue]{hyperref}
\usepackage[tight, FIGTOPCAP, hang, raggedright, nooneline]{subfigure}
\DeclareMathOperator{\tr}{tr}
\newcommand\ket [1] {|#1 \rangle }

\newcommand{\av}[1]{\langle #1\rangle}

\begin{document}

\title{From fractional Chern insulators to Abelian and non-Abelian fractional quantum Hall states: Adiabatic continuity and orbital entanglement spectrum}

\author{Zhao Liu}
\email{zliu@csrc.ac.cn}
\affiliation{Beijing Computational Science Research Center, Beijing, 100084, China}
\affiliation{Institute of Physics, Chinese Academy of Sciences, Beijing, 100190, China}
\author{Emil J. Bergholtz}
\email{ejb@physik.fu-berlin.de}
\affiliation{Dahlem Center for Complex Quantum Systems and Institut f\"ur Theoretische Physik, Freie Universit\"at Berlin, Arnimallee 14, 14195 Berlin, Germany}

\date{\today}

\begin{abstract}
The possibility of realizing lattice analogs of fractional quantum Hall (FQH) states, so-called fractional Chern insulators (FCIs), in nearly flat topological (Chern) bands has attracted a lot of recent interest. Here, we make the connection between Abelian as well as non-Abelian FQH states and FCIs more precise. Using a gauge-fixed version of Qi's Wannier basis representation of a Chern band, we demonstrate that the interpolation between several FCI states, obtained by short-range lattice interactions in a spin-orbit-coupled kagome lattice model, and the corresponding continuum FQH states is smooth: the gap remains approximately constant and extrapolates to a finite value in the thermodynamic limit, while the low-lying part of the orbital entanglement spectrum remains qualitatively unaltered. The orbital entanglement spectra also provide a first glimpse of the edge physics of FCIs via the bulk-boundary correspondence. Corroborating these results, we find that the squared overlaps between the FCI and FQH ground states are as large as $98.7\%$ for the 8-electron Laughlin state at $\nu=\frac{1}{3}$ (consistent with an earlier study) 
and $97.8\%$ for the 10-electron Moore-Read state at $\nu=\frac{1}{2}$. For the bosonic analogs of these states, the adiabatic continuity is also shown to hold, albeit with somewhat smaller associated overlaps, etc. Although going between the Chern bands to the Landau level problem is often smooth, we show that this is not always the case by considering fermions at filling fraction $\nu=\frac{4}{5}$, where the interpolation between Hamiltonians describing the two systems results in a phase transition.
\end{abstract}

\maketitle

\section{Introduction}
After Haldane's seminal work modeling an integer quantum Hall (IQH) effect in a simple lattice model, \cite{haldanemodel} it took over 20 years until it was recently realized that similar ideas were used to emulate lattice analogs of fractional quantum Hall (FQH) states.\cite{chernins1,chernins2,chernins3,cherninsnum1,cherninsnum2} These states, termed fractional Chern insulators (FCIs), have a number of appealing traits: most saliently they do not require an external magnetic field and they might, in principle, persist at elevated temperatures.

While the basic ingredient needed for the IQH effect is a band with non-zero Chern number (and a finite band gap), an additional prerequisite for the FCIs is that these bands are only weakly dispersive, thus enhancing the effect of interactions within the band.
Following the initial suggestions,\cite{chernins1,chernins2,chernins3} there are by now many known models with nearly flat bands carrying non-zero Chern number, including intriguing solid-state proposals\cite{chernins1,c1a,c1c,c1d,c1e,c2,max} and possible cold-atom realizations.\cite{norman} Although there is plenty of numerical evidence for FCI analogs of Laughlin states,\cite{chernins3,cherninsnum1,cherninsnum2} hierarchy/composite fermion states,\cite{beyondL,andreas,ifwlong} as well as non-Abelian states,\cite{nonab1,nonab2,nonab3} the physics in Chern bands is only identical to that of a Landau level in a very idealized limit.\cite{bands,nonab1}
In actual lattice models, however, the distinctions are rather striking such as particle-hole symmetry breaking\cite{andreas,ifwlong} and the emergence of qualitatively new competing compressible states.\cite{andreas} underscoring the need for a better understanding of theses systems at a quantitative level.

In an insightful paper, Qi introduced a Wannier basis representation of a Chern band mincing the Landau gauge wave functions in the continuum, and thereby paved the way towards a more direct comparison between FCI and FQH states.\cite{qi} Indeed, Scaffidi and M\"oller recently used this mapping to convincingly show that the $\nu=\frac{1}{2}$ bosonic FCI state on the honeycomb lattice is indeed smoothly connected to the Laughlin state describing the continuum FQH state at the same filling fraction.\cite{moller} However, a direct implementation of Qi's Wannier mapping is not always successful, e.g., wave-function overlaps with FQH model states often turn out to be minuscule even in models where there are well-established FCI phases, due to the finite-size properties (non-orthogonality) of the Wannier functions and, in principle, also because the two systems carry independent gauge degrees of freedom. This issue was considered in detail by Wu {\it et. al.} who also came up with an involved, yet elegant, prescription that remedies these problems and showed that it leads to impressive overlaps between the fermionic FCI at $\nu=\frac{1}{3}$ and the corresponding Laughlin FQH state.\cite{gaugefixing}

In this work, we apply the Wannier state mapping\cite{qi} adopted to finite-size systems\cite{gaugefixing} to both Abelian and non-Abelian FCI phases. We demonstrate the adiabatic continuity between these states and their corresponding FQH analogs (Laughlin\cite{laughlin83} and Moore-Read\cite{mr} states) by showing that the gap remains essentially unaltered when interpolating between the FCI and FQH Hamiltonians as well as studying the overlaps with the model FQH states which turn out to remain high throughout the interpolation. Moreover, we report on the first studies of orbital entanglement spectra\cite{LiH} (OES) of the FCI states. In contrast to the earlier particle entanglement spectrum\cite{PES} (PES) studies\cite{cherninsnum2} which probe the quasi-hole physics, our OES studies, based on a cut in (Wannier) orbital space,\cite{orbitalcut} provides a test of the edge physics in FCI phases. The upshot of these studies is that the FCI states considered here are, in a well-defined sense, closer to the idealized model FQH wave functions than FQH states obtained for more realistic (Coulomb) interactions in continuum Landau level. Underscoring that these results are indeed non-trivial, we also provide an example where the interpolation between the Landau-level physics and the interacting Chern band problem is not smooth by considering fermions at $\nu=\frac{4}{5}$.

The remainder of this work is organized as follows. In Sec. \ref{model} we give relatively detailed description of the Wannier mapping providing a bridge between the description of Chern bands and continuum Landau levels on a torus. Section \ref{cont} contains our main results on the adiabatic continuity and the OES studies focusing on electronic (fermionic) states (corresponding results for bosons are contained in the Appendix). Finally, we discuss our findings in Sec. \ref{discussion}.

\section{Model and methods}\label{model}
In this section, we put the description of fractional quantum Hall systems in the continuum and
fractional Chern insulators in the lattice on the same footing. First, we discuss the lowest Landau level (LLL) on a torus,\cite{Haldane85PRL} and then we go on to discuss a suitably adapted version of the Wannier function mapping of Chern bands in a finite-size system.\cite{gaugefixing} This provides the necessary framework for a direct quantitative comparison between FCIs with FQH states despite the fact that the two systems have different symmetries. Finally, we give a specific kagome lattice model that we use throughout this work to study the FQH-FCI correspondence.

\subsection{Quantum Hall states}
We consider $N$ particles projected to the lowest Landau level on a twisted
torus spanned by two basic vectors $\textbf{L}_1=L_1\textbf{v}_1(\alpha)$ and $\textbf{L}_2=L_2\textbf{v}_2$,
where $\textbf{v}_1(\alpha)=\sin\alpha\textbf{e}_x+\cos\alpha\textbf{e}_y$,
$\textbf{v}_2=\textbf{e}_y$, where $\alpha$ is the twisted angle of the torus, and $L_{1(2)}$ is the length
of the basic vector (in units of the magnetic length). Assuming the number of flux quanta, $N_s$, through surface of the torus is
an integer, the magnetic translation invariance in the $\textbf{v}_1$ and $\textbf{v}_2$
directions leads to $L_1L_2 \sin\alpha=2\pi N_s$. There are precisely $N_s$ single-particle states, $\ket{\psi_j}$, in the lowest Landau level that we choose as maximally to be localized in the $\textbf{e}_x$ direction
(but delocalized in the $\textbf{e}_y$ direction) as
\begin{eqnarray}
\langle x,y|\psi_j\rangle=\Big(\frac{1}{\sqrt{\pi}L_2}\Big)^{\frac{1}{2}}\sum_{n=-\infty}^{+\infty}
\textrm{exp}\Big\{\textrm{i}\Big(\frac{2\pi j}{L_2}+nL_1\sin\alpha\Big)
\Big(y\nonumber\\
-\frac{2\pi j}{L_2}\cot\alpha-nL_1\cos\alpha\Big)
-\frac{1}{2}\Big(x-\frac{2\pi j}{L_2}-nL_1\sin\alpha\Big)^2\Big\},\nonumber\\
\label{psij}
\end{eqnarray}
where $j=0,1,2,...,N_{s}-1$ is the single-particle
momentum in units of $2\pi /L_2$. Note that $\psi_{j}$ is quasi-periodic and centered
along the line $x=2\pi j/N_2$. We define $N_0$ is the greatest common divisor of $N$ and $N_s$, namely $N_0\equiv\textrm{GCD}(N,N_s)$. Then $p\equiv N/N_0$ and $q\equiv N_s/N_0$ are coprime.
There are two translation
operators, $T_\alpha(\alpha =1,2)$, that commute with the many-body
Hamiltonian (as with any translational invariant operator) and
obey $T_1T_2=e^{2\pi \textrm{i} p/q}T_2T_1$. $T_1$
corresponds to a $\textbf{e}_y$ translation and $T_2$
translates a many-body state one lattice constant $2\pi/L_2$
in the $\textbf{e}_x$ direction. At filling factor $\nu=p/q$, because $T_2^q$ commutes with $T_1$,
we can diagonalize certain many-body Hamiltonian $H_{\textrm{FQH}}$ in the LLL orbital basis
and obtain the many-body ground states $|\Psi_\textrm{FQH}(K_1,K_2)\rangle$ as the common eigenstates of $T_1$ and $T_2^q$ with eigenvalues $e^{2\pi \textrm{i} K_1/N_s}$ and $e^{2\pi \textrm{i} K_2/N_0}$, where
$K_1$ can be regarded as the total momentum in the $\textbf{e}_x$ direction.
It directly follows that the degeneracy of $|\Psi_\textrm{FQH}(K_1,K_2)\rangle$ is at least $q$-fold,
among which we can always pick up $q$-fold center-of-mass degenerate states with
different $K_1$ that are connected by the operator $T_2^k (k=0,1,...,q-1)$.

\begin{figure}
\centerline{\includegraphics[width=\linewidth]{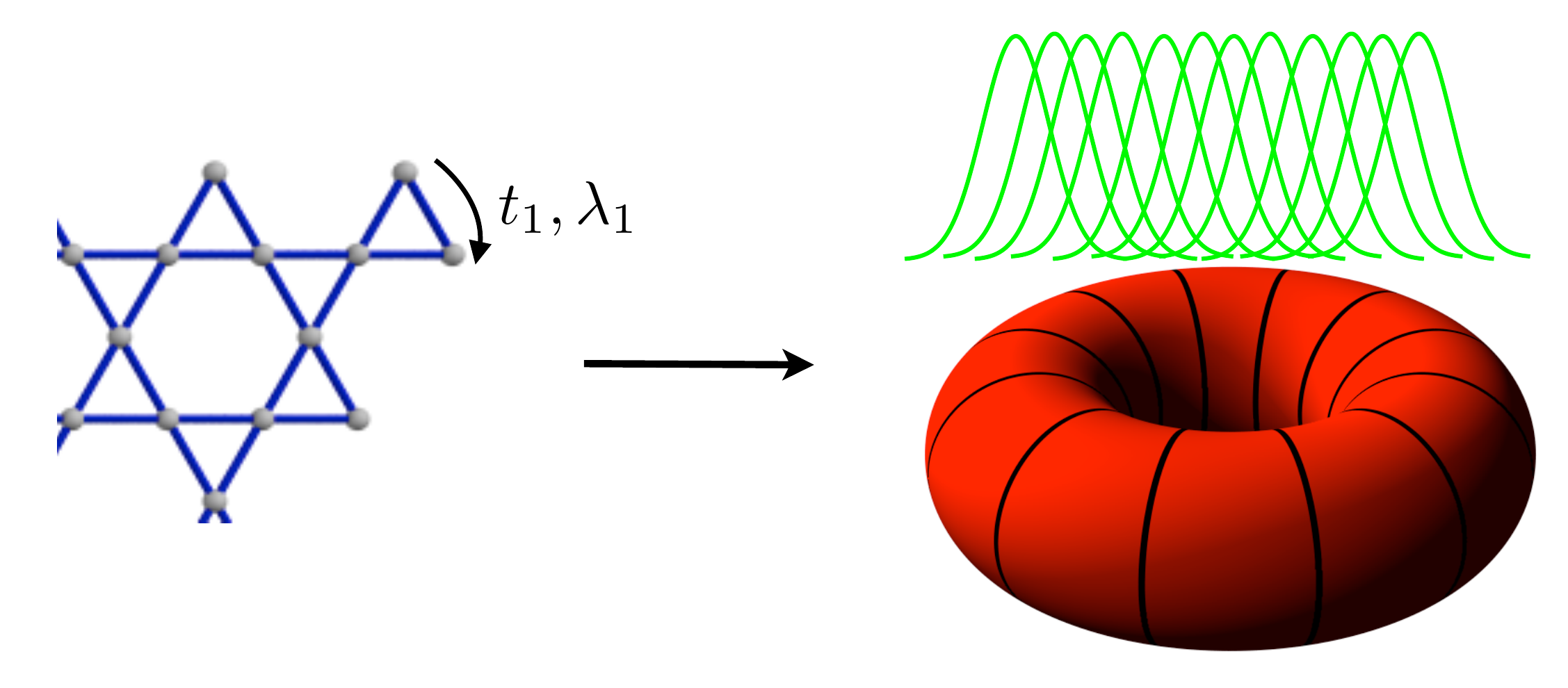}}
\caption{(Color online) Schematic picture of the mapping of a flat Chern band in the lattice model to a continuum Landau level in terms of Wannier states.
\label{fcitofqh}}
\end{figure}

For later convenience, we also introduce an alternative description of the translational symmetry on the torus.
Suppose $N_s$ has two factors $N_1$ and $N_2$, namely $N_s=N_1\times N_2$.
After defining $N_{0,1}\equiv\textrm{GCD}(N,N_1)$ and $q_1\equiv N_1/N_{0,1}$,
we can introduce two translation operators $S_1=(T_2)^{q/q_1}$
and $R_2=T_1^{q_1}$. Because $S_1$ commutes with $R_2$,
we can make the many-body ground states as their common eigenstates.
Within this description of the translational symmetry, the $q$-fold center-of-mass degenerate states
are
\begin{eqnarray}
|\Psi_{\textrm{FQH}}(s,r)\rangle=\frac{1}{\sqrt{q_1}}\sum_{m=0}^{q_1-1}
e^{2\pi \textrm{i}m(\frac{sN-\kappa_2}{N_0q_1})}S_1^m T_2^r |\Psi_\textrm{FQH}(K_1,K_2)\rangle,\nonumber\\
\label{psisr}
\end{eqnarray}
where $s=0,1,...,q_1-1$, $r=0,1,...,q/q_1-1$.
If we choose $N_1$ and $N_2$ appropriately, we can make $q_1=1$. Then
$|\Psi_{\textrm{FQH}}(s,r)\rangle$ and $|\Psi_\textrm{FQH}(K_1,K_2)\rangle$
reduce to the same description.

\subsection{Chern insulators}
Now we move our attention from FQH states in the continuum to the FCIs in the lattice.
We consider a two-dimensional (2D) lattice on the torus with two lattice vectors
$\textbf{v}_1(\beta)=\sin\beta\textbf{e}_x+\cos\beta\textbf{e}_y$ and $\textbf{v}_2=\textbf{e}_y$.
The number of unit cells is $N_1$ and $N_2$ in respective direction and there are $s$ sites in each unit cell.
The states in the first Brillouin zone (1BZ) can be labeled by a 2D momentum $\textbf{k}=(k_1,k_2)$ where $k_i=0,1,...N_i-1$.
In momentum space, the single-particle Hamiltonian can be written as
$H=\sum_{\textbf{k}\in \textrm{BZ}}(c_{\textbf{k},1}^{\dagger},...,c_{\textbf{k},s}^{\dagger})h(\textbf{k})
(c_{\textbf{k},1},...,c_{\textbf{k},s})^T$ and a band structure is formed.
We focus on a single, isolated band $|\textbf{k}\rangle=\sum_{\alpha=1}^s u_{\alpha}(\textbf{k})
c_{\textbf{k},\alpha}^{\dagger}|\textrm{vac}\rangle$, where $u_{\alpha}(\textbf{k})$ is
the corresponding eigenfunction of $h(\textbf{k})$, and suppose $N$ interacting particles
fractionally fill in this band. If the interaction Hamiltonian $H_\textrm{\textrm{FCI}}$
is chosen appropriately, the ground states of this interacting many-body system
are FCI states $|\Psi_{\textrm{FCI}}\rangle$ at certain filling factors $\nu=N/(N_1N_2)$.

To compare the FCI states with the theoretically much better understood FQH states, we need to expand $|\Psi_\textrm{FCI}\rangle$ in a basis with single-particle states that mimic the LLL states [Eq. (\ref{psij})].
An appropriate choice is the Wannier basis, the single-particle state $|X,k_2\rangle$ of which is localized in the $\textbf{v}_1$
direction but delocalized in the $\textbf{v}_2$
direction, where $X$ is the position in the $\textbf{v}_1$
direction and $k_2$ is the momentum (in units of $2\pi/N_2$) in the $\textbf{v}_2$
direction.\cite{qi,gaugefixing}

In a $N_1\times N_2$ finite-size lattice, when focusing on one fractionally filled band
with Chern number $\mathcal{C}$, the (lattice version of the) Berry connection in the $\textbf{v}_1$ direction
can be defined as $\mathcal{A}_1(k_1,k_2)=\sum_\alpha e^{-\textrm{i}2\pi\epsilon_\alpha^1/N_1}
u^*_\alpha(k_1,k_2)u_\alpha(k_1+1,k_2)$, where $\epsilon_\alpha^1$ is the $\textbf{v}_1$ direction relative
displacement of site $\alpha$ in an unit cell. Similarly, we can define the Berry connection
in the $\textbf{v}_2$ direction as $\mathcal{A}_2(k_1,k_2)=\sum_\alpha e^{-\textrm{i}2\pi\epsilon_\alpha^2/N_2}
u^*_\alpha(k_1,k_2)u_\alpha(k_1,k_2+1)$. To restore the orthogonality between different Wannier functions,
we need to introduce unitary Berry connections\cite{gaugefixing}
$A_{1(2)}(k_1,k_2)=\mathcal{A}_{1(2)}(k_1,k_2)/|\mathcal{A}_{1(2)}(k_1,k_2)|$.
Then, the unitary Wilson loops are $W_1(k_2)=\prod_{k_1=0}^{N_1-1}A_1(k_1,k_2)$
and $W_2(k_1)=\prod_{k_2=0}^{N_2-1}A_2(k_1,k_2)$, whose argument angles
are picked in $(-2\pi,0]$.
After defining a shift $\delta_2$ as the cardinality of the set
$\{k_2=0,1,...,N_2|\arg[W_1(k_2)]>\arg[W_1(0)]\}$, we can introduce
a principal Brillouin zone (pBZ) as the set of $k_2$ satisfying $\mathcal{C}k_2+\delta_2\in[0,N_2)$
and move $k_2$ from 1BZ to pBZ.

After introducing $[\lambda_1(k_2)]^{N_1}\equiv W_1(k_2)$
and $[\lambda_2(k_1)]^{N_2}\equiv W_2(k_1)$, where we choose the argument angle
$\arg[\lambda_1(k_2)]\in(-2\pi/N_1,0]$ and $\arg[\lambda_2(k_1)]\in(-2\pi/N_2,0]$,
we can define the Wannier function localized in the $\textbf{v}_1$
direction as
\begin{eqnarray}
|X,k_2\rangle=\frac{e^{\textrm{i}\Phi(k_2)}}{\sqrt{N_1}}
\sum_{k_1=0}^{N_1-1}e^{-\textrm{i}\frac{2\pi k_1}{N_1}X}
\Bigg\{\frac{[\lambda_1(k_2)]^{k_1}}{\prod_{\kappa=0}^{k_1-1}A_1(\kappa,k_2)}\Bigg\}|k_1,k_2\rangle,\nonumber
\label{wannier}
\end{eqnarray}
where $k_2$ is in pBZ and $\Phi(k_2)$ is independent of $X$ and needs to be fixed
by a special prescription (see Appendix \ref{gaugedetails} for details).
Letting $j^{X,k_2}=N_2X+\mathcal{C}k_2+\delta_2$, we can build a one-to-one map between
$|X,k_2\rangle$ and $|\psi_j\rangle$.

Considering the one-to-one map between the Wannier orbital and the LLL orbital as well as their similar localizing
properties, the FCI states $|\Psi_{\textrm{FCI}}\rangle$ in the Wannier basis will be very well
approximated by the lattice version of FQH states constructed as\cite{q_1=1}
\begin{eqnarray}
|\Psi_{\textrm{FQH}}^{\textrm{lat}}(s,r)\rangle=\sum_{\{X,k_2\}}|\{X,k_2\}\rangle\langle\{j^{X,k_2}\}|\Psi_{\textrm{FQH}}(s,r)\rangle,
\label{wanniermap}
\end{eqnarray}
where $|\{\cdots\}\rangle$ is the many-body occupation configuration over the single-particle state $|\cdots\rangle$
(one can find that $|\Psi_{\textrm{FQH}}(s,r)\rangle$ in the LLL orbital basis and its lattice version $|\Psi_{\textrm{FQH}}^{\textrm{lat}}(s,r)\rangle$ in the Wannier basis have a common description). However, it is important to note that $|\Psi_{\textrm{FQH}}^{\textrm{lat}}(s,r)\rangle$
will in general differ from $|\Psi_{\textrm{FCI}}\rangle$, since the Hamiltonians of the two systems have vastly different origins. Moreover, as discussed below, the symmetries of the two models are different.

\subsection{Symmetries}
The FQH Hamiltonian in the LLL on the torus conserves center-of-mass position corresponding to momentum $K_1=\sum_{i=1}^N j_i$ (mod $N_s$). However, the corresponding quantity is not conserved for the FCI problem despite the fact that there is a one-to-one correspondence $j^{X,k_2}=N_2X+\mathcal{C}k_2+\delta_2$ between
$|X,k_2\rangle$ and $|\psi_j\rangle$ which allows us to calculate a total 1D momentum $\sum_{i=1}^N (j^{X,k_2})_i$ (mod $N_1N_2$). Instead, the translational symmetry (in real-space) in the directions of the two lattice vectors in the Chern band implies a conserved two-dimensional momentum, which leads to a reduced symmetry for the FCI Hamiltonian in the Wannier basis: only $J_1=\sum_{i=1}^N (j^{X,k_2})_i$ (mod $N_2$) is conserved. (Another manifestation of the lower symmetry in the FCI problem is reflected in the lack of particle-hole symmetry.\cite{andreas})

The symmetry difference is indeed a generic effect due to the underlying lattice where the Berry curvature necessarily varies in reciprocal space as long as the number of bands is finite. In an ideal limit, however, the FCI Hamiltonian will have the same emergent symmetries as the FQH Hamiltonian .\cite{bands}

\subsection{Kagome lattice model}
In the following, we focus on a special lattice model, namely the kagome lattice model
proposed in Ref. \onlinecite{chernins1}, to investigate the FCI-FQH correspondence.
The single-particle Hamiltonian of the kagome lattice model (cf. Fig. \ref{fcitofqh}) in the real space is
\begin{eqnarray}
H=t_{1}\sum_{\av{i,j},\sigma}c_{i\sigma}^{\dagger}c_{j\sigma}+\textrm{i}\lambda_{1}\sum_{\av{i,j},\alpha,\beta}
(\hat{\mathbf{E}}_{ij}\times \hat{\mathbf{R}}_{ij})\cdot\mathbf{\sigma}_{\alpha\beta}c_{i\alpha}^{\dagger}c_{j\beta},
\nonumber
 \label{1pham}
\end{eqnarray}
where $\hat{\mathbf{E}}_{ij}$ is the normalized,  $|\hat{\mathbf{E}}_{ij}|=1$, electric field arising from an ion at the center of each hexagon as experienced by a particle hopping along the unit-vector $\hat{\mathbf{R}}_{ij}$ from site $i$ to site $j$.
In this work we consider electrons are spin-polarized (all spin up) particles and set $t_1=-1$ while using band structures corresponding to various $\lambda_1$ as input to our studies of interactions projected to non-trivial bands. In the momentum space we have three energy bands, the lowest one of which has Chern number $\mathcal{C}=1$.
As customary,\cite{cherninsnum2} we take the flat band limit and project the interaction Hamiltonian $H_{\textrm{FCI}}$ to this $\mathcal{C}=1$ band. Recent numerical work has indeed shown that both Abelian and non-Abelian FCI states exist in this model.\cite{nonab3}

\section{Continuity between FQH and FCI} \label{cont}

As discussed above, the FQH Hamiltonian and FCI Hamiltonian have
different symmetries. However, they have similar expression written in second-quantized form. Taking two-body interactions as an example, we have $H_{\textrm{FQH}}=\sum_{j_1j_2j_3j_4=0}^{N_s-1}\delta_{j_1+j_2,j_3+j_4}^{\textrm{mod} N_s}
V_{j_1j_2j_3j_4}^{\textrm{FQH}}c_{j_1}^\dagger c_{j_2}^\dagger c_{j_3} c_{j_4}$, for the FQH case, where
$c_j^\dagger$ ($c_j$) creates (annihilates) a particle in the state $|\psi_j\rangle$, while we have $H_{\textrm{FCI}}=\sum_{j_1j_2j_3j_4=0}^{N_1N_2-1}\delta_{j_1+j_2,j_3+j_4}^{\textrm{mod} N_2}
V_{j_1j_2j_3j_4}^{\textrm{FCI}}c_{j_1}^\dagger c_{j_2}^\dagger c_{j_3} c_{j_4}$, where
$c_j^\dagger$ ($c_j$) creates (annihilates) a particle in the state $|X,k_2\rangle$ with
$j=XN_2+\mathcal{C}k_2+\delta_2$ for the FCI case.
Therefore, the structure of the Hilbert space
of FCIs in the Wannier basis is the same as that of FQH systems in the LLL orbital basis if we set $N_s=N_1\times N_2$.
This makes it meaningful to consider an interpolating Hamiltonian as follows,
\begin{eqnarray}
H(\lambda)=\lambda w_{\textrm{FCI}}H_{\textrm{FCI}}+(1-\lambda)w_{\textrm{FQH}}H_{\textrm{FQH}},
\label{interpolate}
\end{eqnarray}
where $\lambda\in[0,1]$ is the interpolation parameter and
$w_{\textrm{FCI}}$ and $w_{\textrm{FQH}}$ are the energy rescaling factors
that can make the energy gap at $\lambda=0$ and $\lambda=1$ equal to 1 (in the cases where we find adiabatic continuity below it is well established that the gap survives in the thermodynamic limit at $\lambda=0$ and $\lambda=1$ respectively).
We then diagonalize Eq. (\ref{interpolate}) in each $J_1=0,1,...,N_2-1$ sector
and analyze the energy gap and the ground states as a function of $\lambda$,
in order to examine whether the FCI states are adiabatically connected to the corresponding FQH model states.

\subsection{Fermions at $\nu=\frac{1}{3}$}
We start our discussion by focusing on the fermions at filling factor $\nu=\frac{1}{3}$.
On the FQH side, we choose the Hamiltonian as $H_{\textrm{FQH}}=\sum_{i<j}\nabla_i^2\delta^2(\textbf{r}_i-\textbf{r}_j)$.
Then the ground states are exact threefold-degenerate fermionic Laughlin states with zero energy. On the FCI side, the Hamiltonian is set as
the nearest-neighbor interaction $H_{\textrm{FCI}}=\sum_{\langle ij \rangle }n_in_j$.
The ground states are three nearly degenerate states separated by a gap
from the excited states. Since $\beta=\pi/3$ for the kagome lattice,
we set $\alpha=\pi/3$ also for the twisted Landau-level torus. This choice is further justified
by the large overlap between the FCI states at $\lambda=1$ and the Laughlin states
at $\lambda=0$ (see the following discussion).

\begin{figure}
\centerline{\includegraphics[height=7.5cm,width=0.8\linewidth]{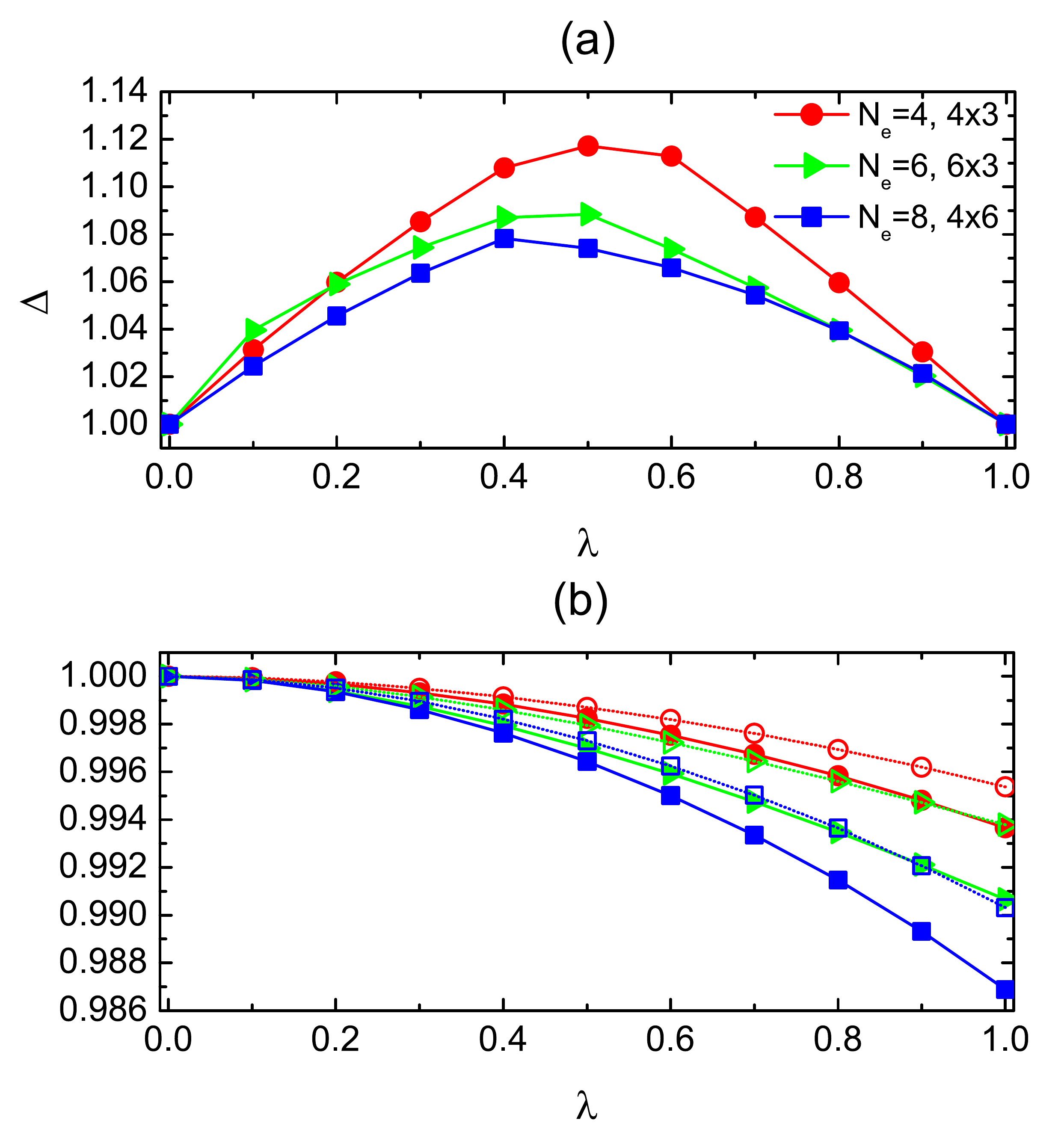}}
\caption{(Color online) Results of the interpolation Eq. (\ref{interpolate})
for fermions at $\nu=\frac{1}{3}$ with $N_e=4$ (red dot), $N_e=6$ (green triangle),
and $N_e=8$ (blue square). In the FCI part, the lattice size is $N_1\times N_2=4\times3$,
$N_1\times N_2=6\times3$ and $N_1\times N_2=4\times6$, respectively; and
$\lambda_1=1$.
(a) The energy gap $\Delta$ does not close for any intermediate $\lambda$.
(b) The total overlap $\mathcal{O}_{\textrm{tot}}$ (filled symbol, solid line) and
the average weight $\overline{\mathcal{W}}$ (empty symbol, dotted line) are still close to 1 at $\lambda=1$.
All of those demonstrate that the continuity holds for fermions at $\nu=\frac{1}{3}$.}
\label{ferlau}
\end{figure}

We find that for each $\lambda\in[0,1]$, there are three nearly-degenerate
states separated by a sizable gap $\Delta$ from excited levels in the energy spectrum. We report the
evolution of $\Delta$ with $\lambda$ for various system sizes in Fig. \ref{ferlau}(a).
It can be seen that the gap never closes for any intermediate $\lambda$; in fact, it is always greater than one and
has a maximal value at $\lambda\approx0.4-0.6$. This provides strong evidence for the
adiabatic continuity between FQH states and FCI states.

To further confirm that
there is no phase transition between $\lambda=0$ and $1$, we study the properties
of the ground manifold. We can define the total overlap as
$\mathcal{O}_{\textrm{tot}}=\frac{1}{d}\sum_{i=1}^d\sum_{j=1}^d|\langle\Psi_{\textrm{FQH}}^i|\Psi^j(\lambda)\rangle|^2$,
where $d$ is the number of (nearly) degenerate states (here, $d=3$),
$|\Psi^j(\lambda)\rangle$ is the (nearly) degenerate state of $H(\lambda)$, and
$|\Psi_{\textrm{FQH}}^j\rangle=|\Psi^j(\lambda=0)\rangle$ is the FQH state (here, the exact $\nu=\frac{1}{3}$ Laughlin state).
We find that $\mathcal{O}_{\textrm{tot}}$ decreases from 1 at $\lambda=0$
smoothly to about 0.987 at $\lambda=1$ for our largest system size $N_e=8$ [Fig. \ref{ferlau}(b)].
Therefore, the ground states do not change
qualitatively during the interpolation from $\lambda=0$ to $\lambda=1$,
supporting that the FCI states are indeed very well captured by the lattice version of FQH states
constructed by Eq. (\ref{wanniermap}).

\begin{figure*}
\centerline{\includegraphics[width=\linewidth]{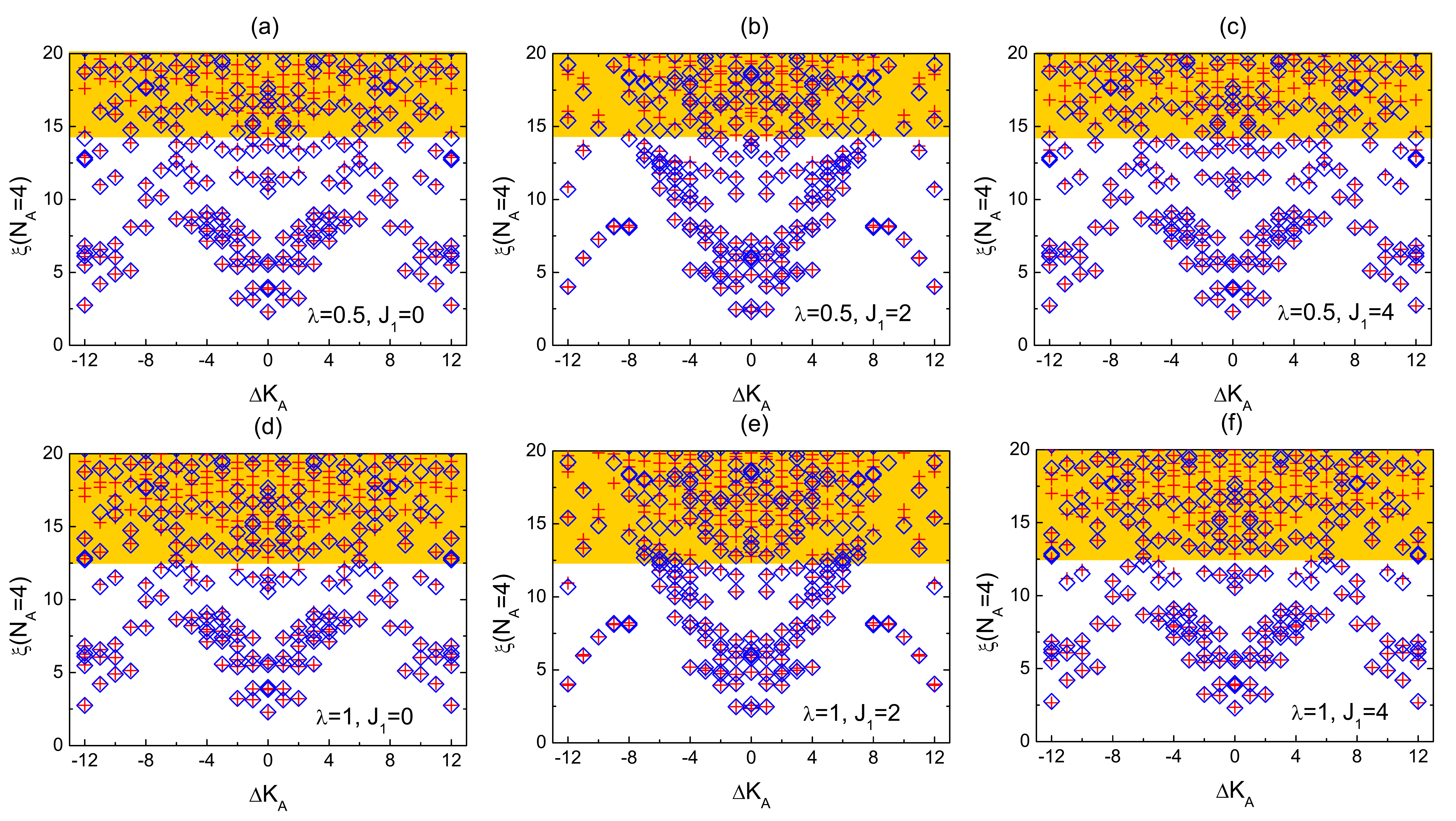}}
\caption{(Color online) The orbital entanglement spectra (OES) of exact fermionic Laughlin states (blue diamond) and the projected nearly-degenerate states $|\Psi_{\textrm{prj}}^{J_1}(\lambda)\rangle$ (red cross) at $\nu=\frac{1}{3}$, $N_e=8$.
The lattice size is $N_1\times N_2=4\times6$ and $\lambda_1=1$ for the FCI part.
In the left column, $J_1=0$,
corresponding to the Laughlin state in $K_1=12$ sector.
In the middle column, $J_1=2$,
corresponding to the Laughlin state in $K_1=20$ sector.
In the right column, $J_1=4$,
corresponding to the Laughlin state in $K_1=4$ sector.
In (a), (b) and (c), $\lambda=0.5$.
(a): The unprojected $|\Psi^{J_1=0}(\lambda)\rangle$ has weight $\mathcal{W}\approx0.99730$ on $K_1=12$ sector.
The overlap with the Laughlin state $\mathcal{O}=|\langle\Psi_{\textrm{Lau}}^{K_1=12}|\Psi^{J_1=0}(\lambda)\rangle|^2$ is 0.99654.
(b): The unprojected $|\Psi^{J_1=2}(\lambda)\rangle$ has weight $\mathcal{W}\approx0.99732$ on $K_1=20$ sector.
The overlap with the Laughlin state $\mathcal{O}=|\langle\Psi_{\textrm{Lau}}^{K_1=20}|\Psi^{J_1=2}(\lambda)\rangle|^2$ is 0.99638.
(c): The unprojected $|\Psi^{J_1=4}(\lambda)\rangle$ has weight $\mathcal{W}\approx0.99732$ on $K_1=4$ sector.
The overlap $\mathcal{O}=|\langle\Psi_{\textrm{Lau}}^{K_1=4}|\Psi^{J_1=4}(\lambda)\rangle|^2$ is 0.99638.
In (d), (e) and (f), $\lambda=1$.
(d): The unprojected $|\Psi^{J_1=0}(\lambda)\rangle$ has weight $\mathcal{W}\approx0.99062$ on $K_1=12$ sector.
The overlap with the Laughlin state $\mathcal{O}=|\langle\Psi_{\textrm{Lau}}^{K_1=12}|\Psi^{J_1=0}(\lambda)\rangle|^2$ is 0.98733.
(e): The unprojected $|\Psi^{J_1=2}(\lambda)\rangle$ has weight $\mathcal{W}\approx0.99034$ on $K_1=20$ sector.
The overlap with the Laughlin state $\mathcal{O}=|\langle\Psi_{\textrm{Lau}}^{K_1=20}|\Psi^{J_1=2}(\lambda)\rangle|^2$ is 0.98667.
(f): The unprojected $|\Psi^{J_1=4}(\lambda)\rangle$ has weight $\mathcal{W}\approx0.99034$ on $K_1=4$ sector.
The overlap with the Laughlin state $\mathcal{O}=|\langle\Psi_{\textrm{Lau}}^{K_1=4}|\Psi^{J_1=4}(\lambda)\rangle|^2$ is 0.98667.
The weight and overlap in the middle column and right column are always the same because of the inversion symmetry
of the Wannier basis.\cite{gaugefixing}
The orange shadows indicate the generic levels in the OES of $|\Psi_{\textrm{prj}}^{J_1}(\lambda)\rangle$
which deviate from the levels of the exact Laughlin state.
\label{ferlauoes}}
\end{figure*}

The entanglement spectrum\cite{LiH} (ES) can usually provide us more insights than the overlap, which is only
a single number and will necessarily vanish in the thermodynamic limit.
For any bipartite pure state $|\Psi\rangle_{AB}$, it can be decomposed using the Schmidt decomposition,
\begin{equation} \label{schmidt}
|\Psi\rangle_{AB}=\sum_i
e^{-\xi_i/2}|\phi_{i}^A\rangle\otimes|\phi_{i}^B\rangle,\nonumber
\end{equation}
where the states $|\phi_{i}^A\rangle$ ($|\phi_{i}^B\rangle$) form an
orthonormal basis for the subsystem $A$ ($B$). $\{\xi_i\geq 0\}$ is defined as
entanglement spectrum and is related to the eigenvalues,
$\eta_i$, of the reduced density matrix,
$\rho_A=\tr_B(|\Psi\rangle_{AB \ AB}\langle\Psi|)$, of $A$ as
$\eta_i=e^{-\xi_i}$.
In some previous works, the ES for particle cut has been investigated
extensively to probe the quasihole excitation properties of FCI states.
Here, we focus on another kind of ES, the OES for a cut in orbital space, to test
the edge physics of FCI states.

We first briefly recall the OES of FQH states on the torus that has been studied in Refs. \onlinecite{Lauchli,Zhao}.
The threefold-degenerate fermionic $\nu=\frac{1}{3}$ Laughlin states have the following simple representations
in the orbital basis in the thin-torus limit\cite{bklong} $L_2=0$ (for $N_e=8$, $N_s=24$):
\begin{eqnarray}
&100100|\textbf{100100100100}|100100,&\nonumber\\
&010010|\textbf{010010010010}|010010,&\nonumber\\
&001001|\textbf{001001001001}|001001.&
\label{lautt}
\end{eqnarray}
We bipartition the system into blocks $A$ and $B$, which consist of $l_{A}$ consecutive
orbits and the remaining $N_{s}-l_{A}$ orbits, respectively [The bold block in Eq. (\ref{lautt}) is our
subsystem $A$]. After extracting the ES from the ground states, we label
every ES level by the particle number $N_{A}=\sum_{j\in A}n_{j}$ and
the total momentum $K_{A}=\sum_{j\in A}jn_{j}$ (mod $N_{s}$) in
block $A$, where $n_j$ is the particle number in the state $|\psi_j\rangle$. In
this work, we concentrate on the case $l_{A}=N_{s}/2$.

In Refs. \onlinecite{Lauchli,Zhao}, it was shown that the resulting OES for the FQH state form towers that can be decomposed into the edge modes of the underlying conformal field theory (CFT). This combination comes about as the natural partition [Eq. (\ref{lautt})] gives a subsystem $A$ with the geometry of a cylinder which has two edges on which gapless edge states with opposite chirality reside. An illuminating recent discussion of the connection between the OES, the CFT describing the edge and matrix product states was given in Ref. \onlinecite{fqhmps}.

Considering that the structure of the Hilbert space does not change during the interpolation,
we can make a cut in the basis and extract the OES of $|\Psi^i(\lambda)\rangle$ by the same
method as that for FQH states.
For pure FCI states, this corresponds to a cut in the localized Wannier orbitals.
However, the total momentum $K_1$ in $|\Psi^i(\lambda)\rangle$ is not a good quantum number
(except at $\lambda=0$).
This means $|\Psi^i(\lambda)\rangle$ may have weight on some $K_1$ that $|\Psi_{\textrm{FQH}}\rangle$
does not have weight on. We can calculate the weight $\mathcal{W}^i$ of each $|\Psi^i(\lambda)\rangle$
on the $K_1$ sectors of FQH states and obtain an average weight $\overline{\mathcal{W}}=\frac{1}{d}\sum_{i=1}^d \mathcal{W}^i$.
From Fig. \ref{ferlau}(b), we can see that the $|\Psi^i(\lambda>0)\rangle$ indeed has
some "momentum leakage" leading to $\overline{\mathcal{W}}<1$. However, even for the pure FCI states
at $\lambda=1$, $K_1$ is also almost conserved ($\overline{\mathcal{W}}\approx0.990$ for $N_e=8$).
Therefore, we project $|\Psi^i(\lambda)\rangle$ into the $K_1$ sector of the corresponding FQH states
and consider the OES of $|\Psi^i_{\textrm{prj}}(\lambda)\rangle$.

In Fig. \ref{ferlauoes}, we display our OES results for $N_e=8$ (the lattice size is $N_1\times N_2=4\times6$
for the FCI part). For this system size, $|\Psi^i(\lambda)\rangle$ are located in the $J_1=0$, $J_1=2$
and $J_1=4$ sectors. They correspond to the Laughlin state with $K_1=12$ (100 sector), $K_1=20$ (010 sector) and
$K_1=4$ (001 sector), respectively. We find that the OES almost perfectly
match that of the corresponding Laughlin states up to $\xi=\xi_{\textrm{max}}$ for all of the three $|\Psi^i(\lambda)\rangle$:
$\xi_{\textrm{max}}\approx13.3$ at $\lambda=0.5$, while it reduces slightly to about 12.3 at $\lambda=1$. While the notion of an entanglement gap\cite{LiH,ronny} cannot be defined as crisply as in geometries with only one edge,\cite{ronny,Bergholtz-N-S}
the impressive match of the OES with the model state nevertheless strongly suggests that the edge excitation properties of the Laughlin states are preserved during the interpolation and furthermore corroborates the adiabatic continuity between FCI states and FQH states at $\nu=\frac{1}{3}$. In Ref. \onlinecite{Lauchli}, the OES of the $\nu=\frac{1}{3}$ Coulomb ground states were investigated and compared to the model states. In this case there was a match of the OES levels with exact $\nu=\frac{1}{3}$ Laughlin state up to $\xi_{\textrm{max}}\approx8$ (the number of levels below this value increase with system size). That we find higher $\xi_{\textrm{max}}$ here indicates that the FCI states in the lattice are actually closer to the Laughlin model states than is the case for the Coulomb FQH ground states.

\subsection{Fermions at $\nu=\frac{1}{2}$}
It is also interesting to investigate whether the adiabatic continuity holds also for
some non-Abelian states. In fact, none of the two previous Wannier basis studies\cite{gaugefixing,moller} considered states in this class. To this end, we turn our attention to the $\nu=\frac{1}{2}$ fermionic Moore-Read phase. To obtain the exact fermionic Moore-Read states in the continuum on the torus,
we choose $H_{\textrm{FQH}}=\sum_{i<j<k}\mathcal{S}_{ijk}\nabla_i^2\nabla_j^4
\delta^2(\textbf{r}_i-\textbf{r}_j)\delta^2(\textbf{r}_j-\textbf{r}_k)$, where $\mathcal{S}_{ijk}$
is the symmetrizing operator. The ground states are exact sixfold-degenerate fermionic Moore-Read states with zero energy.
On the FCI side, we construct the Hamiltonian as a three-body interaction $H_{\textrm{FCI}}=\sum_{\langle ijk \rangle }n_in_jn_k$
between three nearest-neighbor sites.
The ground states are six nearly degenerate states separated by a gap
from the excited states. Here we choose a different twisted angle $\alpha=2\pi/3$
for the torus and this is justified by the large overlap between the FCI states at $\lambda=1$
and the Moore-Read states at $\lambda=0$ (for $\alpha=\pi/3$ this overlap is relatively small).

\begin{figure}
\centerline{\includegraphics[height=7.5cm,width=0.8\linewidth]{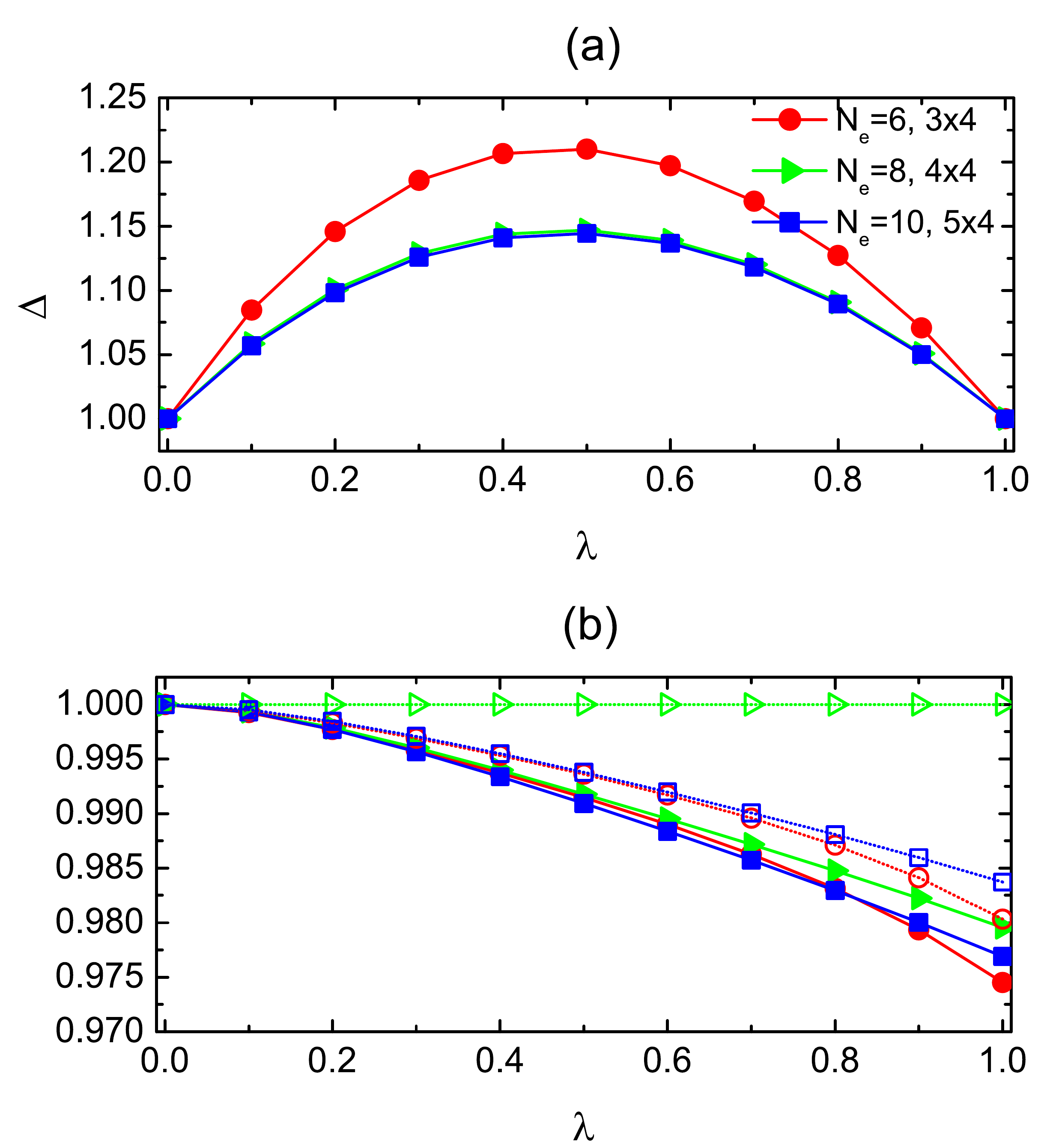}}
\caption{(Color online) Results of the interpolation Eq. (\ref{interpolate}) for
fermions at $\nu=\frac{1}{2}$ for $N_e=6$ (red dot), $N_e=8$ (green triangle),
and $N_e=10$ (blue square). In the FCI part, the lattice size is $N_1\times N_2=3\times4$,
$N_1\times N_2=4\times4$ and $N_1\times N_2=5\times4$, respectively; and
$\lambda_1=0.8$.
(a) The energy gap $\Delta$ does not close for any intermediate $\lambda$.
(b) The total overlap $\mathcal{O}_{\textrm{tot}}$ (filled symbol, solid line) and
the average weight $\overline{\mathcal{W}}$ (empty symbol, dotted line) are still close to 1 at $\lambda=1$.
All of those demonstrate that the continuity holds for fermions at $\nu=\frac{1}{2}$.
(One may note that for $N_e=8$, $N_1\times N_2=4\times4$, $\overline{\mathcal{W}}=1$ for all $\lambda$.
This is accidental for this particular lattice size.)}
\label{ferpf}
\end{figure}

We find that there are six nearly degenerate states $|\Psi^i(\lambda)\rangle$ for each
$\lambda\in[0,1]$. The evolution of the energy gap [Fig. \ref{ferpf}(a)], total overlap, and average weight
[Fig. \ref{ferpf}(b)] behave similarly with those for the $\nu=\frac{1}{3}$ fermionic Laughlin phase.
While both of the total overlap and average weight are slightly smaller ($\mathcal{O}_{\textrm{tot}}\approx0.977$
and $\overline{\mathcal{W}}\approx0.984$ at $\lambda=1$ for our largest system size $N_e=10$), these numbers are way above the overlaps found between the Moore-Read state and the Coulomb ground state in the second Landau level.\cite{rrhaldane} (Of course, the three-body lattice Hamiltonian used here for the FCI is somewhat artificial to begin with making a direct comparison of overlaps a bit biased.)

\begin{figure*}
\centerline{\includegraphics[height=10.5cm,width=0.7\linewidth]{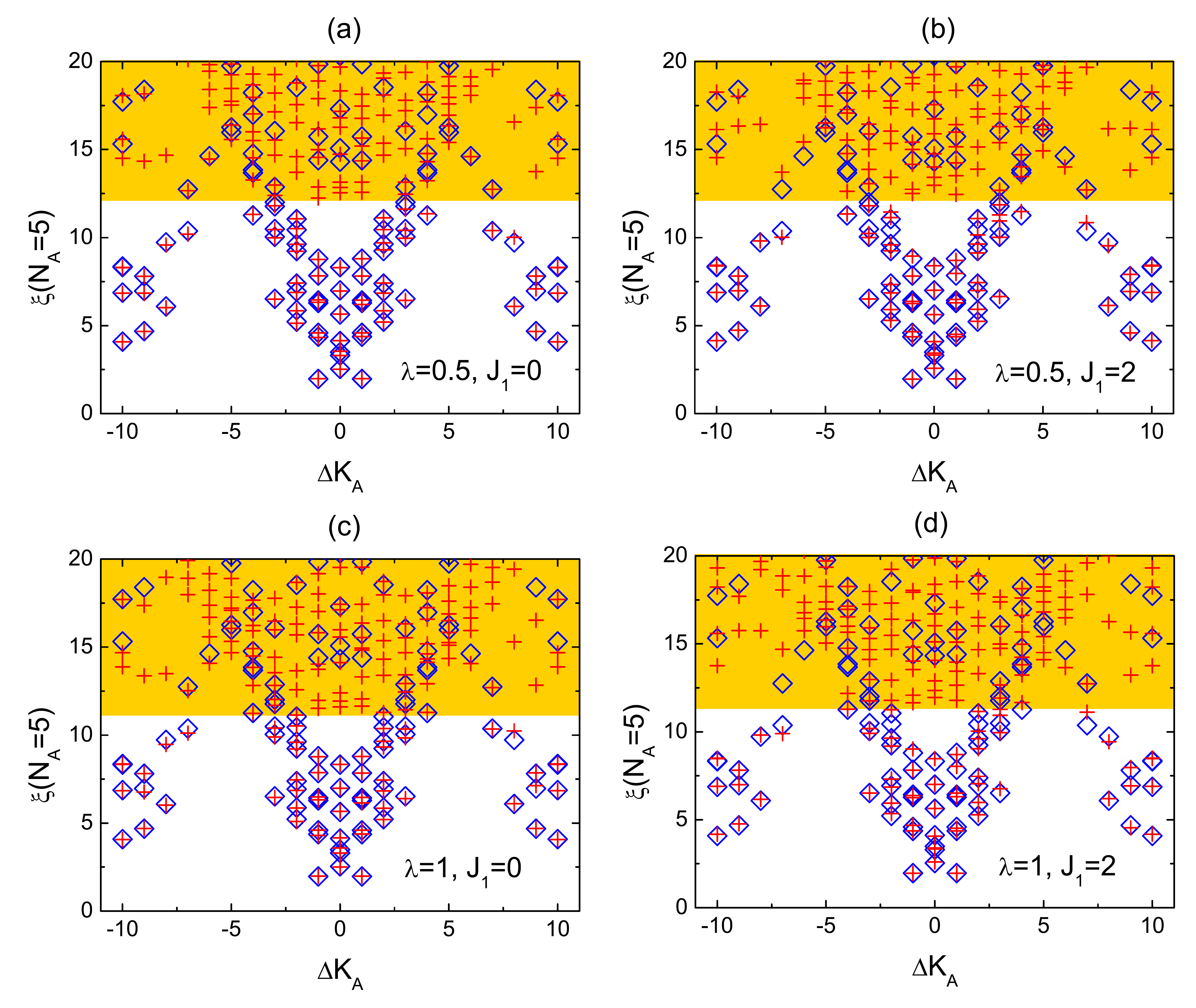}}
\caption{(Color online) The orbital entanglement spectra (OES) of exact fermionic Moore-Read states (blue diamond) and
the projected nearly-degenerate states $|\Psi_{\textrm{prj}}^{J_1}(\lambda)\rangle$ (red cross)
at $\nu=\frac{1}{2}$, $N_e=10$. The lattice size is $N_1\times N_2=5\times4$ and $\lambda_1=0.8$ for the FCI part.
In the left column, $J_1=0$,
corresponding to the Moore-Read state in $K_1=0$ sector.
In the right column, $J_1=2$,
corresponding to the Moore-Read state in $K_1=10$ sector.
In (a) and (b), $\lambda=0.5$.
(a): The unprojected $|\Psi^{J_1=0}(\lambda)\rangle$ has weight $\mathcal{W}\approx0.99403$ on $K_1=0$ sector.
The overlap with the Moore-Read state $\mathcal{O}=|\langle\Psi_{\textrm{MR}}^{K_1=0}|\Psi^{J_1=0}(\lambda)\rangle|^2$ is 0.99119.
(b): The unprojected $|\Psi^{J_1=2}(\lambda)\rangle$ has weight $\mathcal{W}\approx0.99436$ on $K_1=10$ sector.
The overlap with the Moore-Read state $\mathcal{O}=|\langle\Psi_{\textrm{MR}}^{K_1=10}|\Psi^{J_1=2}(\lambda)\rangle|^2$ is 0.99094.
In (c) and (d), $\lambda=1$.
(c): The unprojected $|\Psi^{J_1=0}(\lambda)\rangle$ has weight $\mathcal{W}\approx0.98488$ on $K_1=0$ sector.
The overlap with the Moore-Read state $\mathcal{O}=|\langle\Psi_{\textrm{MR}}^{K_1=0}|\Psi^{J_1=0}(\lambda)\rangle|^2$ is 0.97792.
(d): The unprojected $|\Psi^{J_1=2}(\lambda)\rangle$ has weight $\mathcal{W}\approx0.98583$ on $K_1=10$ sector.
The overlap with the Moore-Read state $\mathcal{O}=|\langle\Psi_{\textrm{MR}}^{K_1=10}|\Psi^{J_1=2}(\lambda)\rangle|^2$ is 0.97750.
The orange shadows indicate the generic levels in the OES of $|\Psi_{\textrm{prj}}^{J_1}(\lambda)\rangle$
which deviate from the levels of the exact Moore-Read state.
\label{ferpfoes}}
\end{figure*}

We also consider the OES for $N_e=10$ (the lattice size is $N_1\times N_2=5\times4$
for the FCI part). In the continuum, the thin-torus configuration of the six fermionic Moore-Read states are\cite{ttpfaff}
(for $N_e=10$, $N_s=20$)
\begin{eqnarray}
&010101|\textbf{0101010101}|0101,&\nonumber\\
&101010|\textbf{1010101010}|1010,&\nonumber\\
&01100|\textbf{1100110011}|00110\pm10011|\textbf{0011001100}|11001,&\nonumber\\
&11001|\textbf{1001100110}|01100\pm00110|\textbf{0110011001}|10011.&\nonumber
\label{pftt}
\end{eqnarray}
Their total momentum is $K_1=0$, $K_1=10$, $K_1=15$ (two-fold) and $K_1=5$ (two-fold), respectively.
Among the six $|\Psi^i(\lambda)\rangle$, one is located in the $J_1=0$ sector [corresponding to the $K_1=0$
Moore-Read state (0101 sector)], one is located in the $J_1=2$ sector [corresponding to the $K_1=10$ Moore-Read
state (1010 sector)], two are located in the $J_1=1$ sector [corresponding to the two $K_1=5$ Moore-Read states
(1100$\pm$0011 sectors)], and two are located in the $K_2=3$ sector [corresponding to the two $K_1=15$ Moore-Read states
(0110$\pm$1001 sectors)]. The two $|\Psi^i(\lambda)\rangle$ with the same $J_1=1$ ($J_1=3$)
will mix with each other, so they do not have a one-to-one correspondence to the 1100+0011 state
and 1100-0011 state (0110+1001 state and 0110-1001 state). Therefore, we only consider the OES for the $|\Psi^i(\lambda)\rangle$ in $J_1=0$
and $J_1=2$ sectors here by projecting them into the $K_1$ sector of the corresponding
Moore-Read states. In Fig. \ref{ferpfoes}, we can see that the low-lying part of the OES of $|\Psi^i_{\textrm{prj}}(\lambda)\rangle$
also match that of the exact Moore-Read state very well. As expected, $\xi_{\textrm{max}}$ here
($\xi_{\textrm{max}}\approx12.5$ at $\lambda=0.5$ and $\xi_{\textrm{max}}\approx11$ at $\lambda=1$)
is lower slightly than that for the $\nu=\frac{1}{3}$ fermionic Laughlin case reflecting the lower overlap.
We also find that the OES of $|\Psi^i_{\textrm{prj}}(\lambda)\rangle$ lacks inversion (left-right) symmetry, which
is not so obvious in the $\nu=\frac{1}{3}$ state. However, taken together there is no doubt that the FCI phase is excellently described by the Moore-Read wave function and the low energy physics of the FCI problem should thus be within the same universality class.

\subsection{Fermions at $\nu=\frac{4}{5}$ and $\nu=\frac{2}{3}$}
\begin{figure}
\centerline{\includegraphics[width=\linewidth]{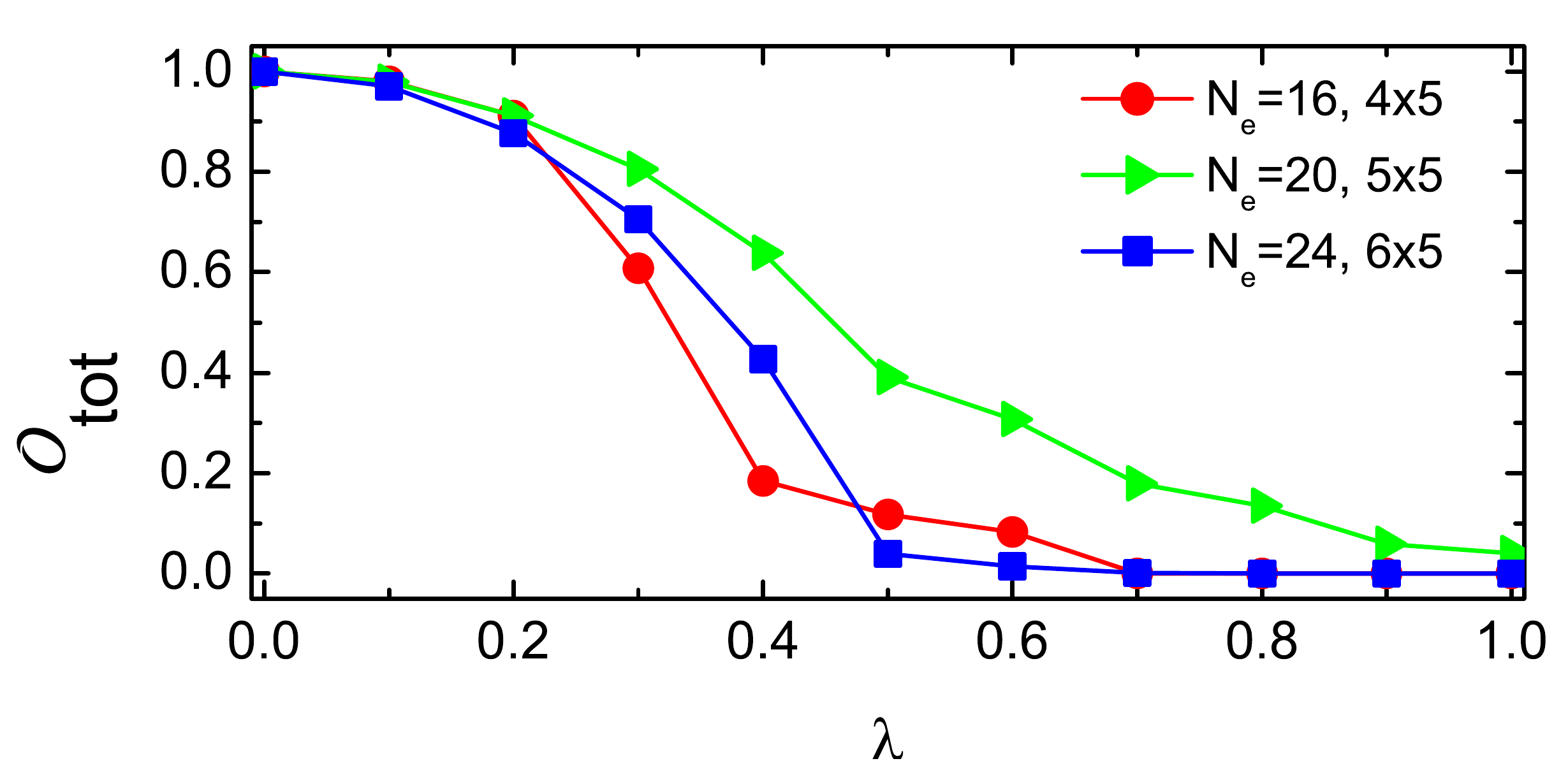}}
\caption{(Color online) Results of the interpolation Eq. (\ref{interpolate}) for fermions at $\nu=\frac{4}{5}$ for $N_e=16$ (red dot),
 $N_e=20$ (green triangle), and $N_e=24$ (blue square). In the FCI part, the lattice size is $N_1\times N_2=4\times5$,
$N_1\times N_2=5\times5$ and $N_1\times N_2=6\times5$, respectively; and
$\lambda_1=1$.
The total overlap $\mathcal{O}_{\textrm{tot}}$ shows a clear drop
at intermediate $\lambda$, which demonstrates that the continuity does not hold for fermions at $\nu=\frac{4}{5}$.
}
\label{fer4over5}
\end{figure}

\begin{figure}
\centerline{\includegraphics[height=10cm,width=0.7\linewidth]{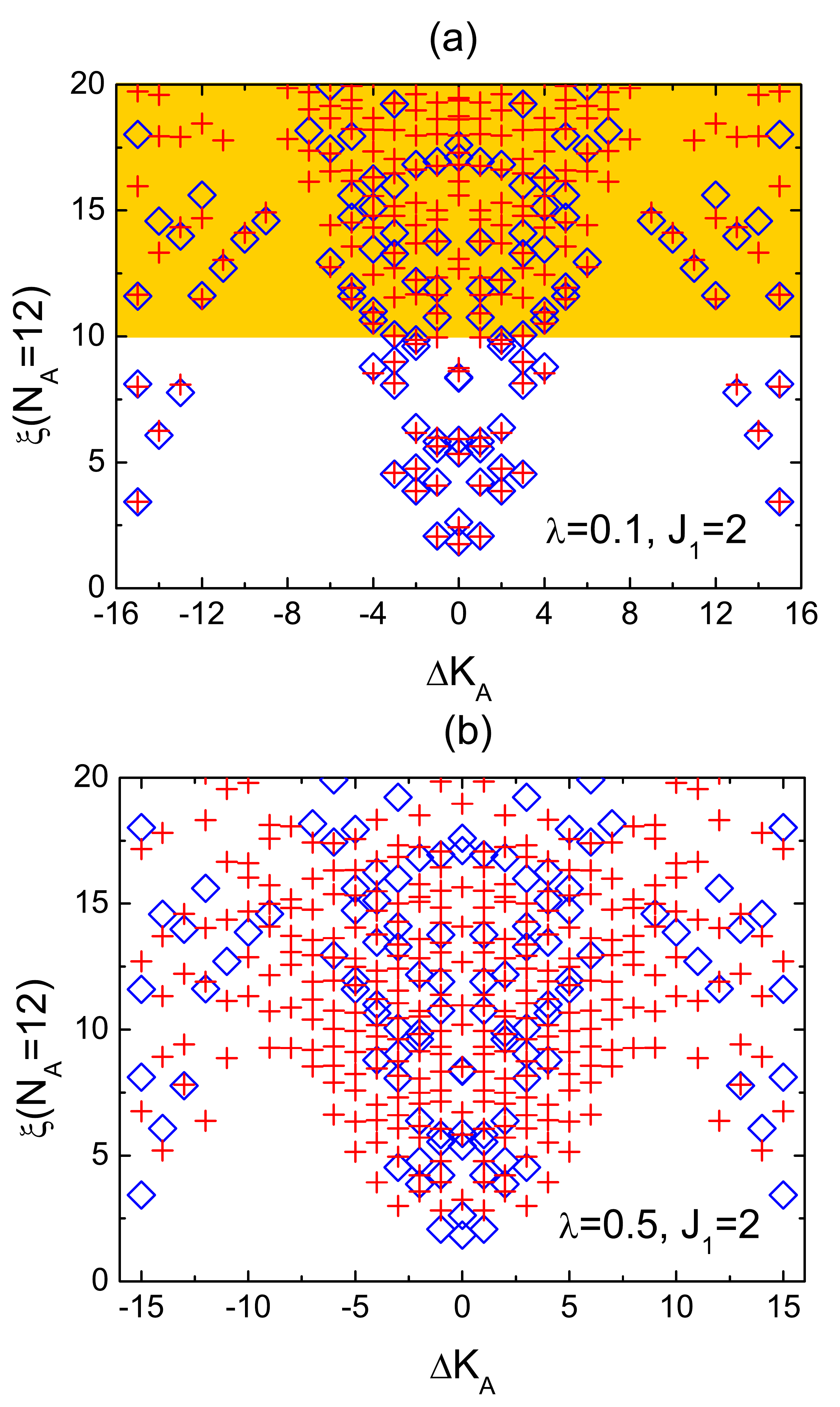}}
\caption{(Color online) The orbital entanglement spectra (OES) of exact fermionic phc state (blue diamond) and
the projected state $|\Psi_{\textrm{prj}}^{J_1}(\lambda)\rangle$ (red cross)
at $\nu=\frac{4}{5}$, $N_e=24$. $|\Psi_{\textrm{prj}}^{J_1}(\lambda)\rangle$ is in $J_1=2$ sector
and corresponds to the phc state in $K_1=12$ sector. The lattice size is $N_1\times N_2=6\times5$ and $\lambda_1=1$ for the FCI part.
(a): $\lambda=0.1$. The unprojected $|\Psi^{J_1}(\lambda)\rangle$ has weight $\mathcal{W}\approx0.98645$ on $K_1=12$ sector.
The overlap with the phc state $\mathcal{O}=|\langle\Psi_{\textrm{phc}}^{K_1=12}|\Psi^{J_1=2}(\lambda)\rangle|^2$ is 0.97213.
(b): $\lambda=0.5$. The unprojected $|\Psi^{J_1}(\lambda)\rangle$ has weight $\mathcal{W}\approx0.06792$ on $K_1=12$ sector.
The overlap with the phc state $\mathcal{O}=|\langle\Psi_{\textrm{phc}}^{K_1=12}|\Psi^{J_1=2}(\lambda)\rangle|^2$ is almost 0.
}
\label{fer4over5oes}
\end{figure}

Finally, we consider a case where there is a lack of adiabatic continuity between the low energy sector of the FQH and FCI Hamiltonians. To this end we focus on fermions at $\nu=\frac{4}{5}$.
On the FQH side, we choose the Hamiltonian as $H_{\textrm{FQH}}=\sum_{i<j}\nabla_i^6\delta^2(\textbf{r}_i-\textbf{r}_j)$.
Then the ground states are fivefold degenerate states that are the particle-hole conjugate (phc) of
$\nu=\frac{1}{5}$ Laughlin states. On the FCI side, the Hamiltonian is set as
the nearest-neighbor interaction $H_{\textrm{FCI}}=\sum_{\langle ij \rangle }n_in_j$.
Due to the particle-hole symmetry breaking, the ground states are no longer FCI states but competing
compressible (fermion-liquid-like) states
without the fivefold nearly degeneracy.\cite{andreas} Here, we set $\beta=\pi/3$, $w_{\textrm{FCI}}=w_{\textrm{FQH}}=1$,
and choose $|\Psi^i(\lambda)\rangle$
as the ground states in the $J_1$ sectors where the FQH states are located in at $\lambda=0$.
We use the total overlap and OES to probe the phase transition between $\lambda=0$ and $\lambda=1$.
In Fig. \ref{fer4over5}, it is clear that the total overlap drops down to a very small number at intermediate $\lambda$.
In Fig. \ref{fer4over5oes}, we choose the $|\Psi^i(\lambda)\rangle$ in the $J_1=2$ sector to
study the OES. One can see that the OES of the projected $|\Psi^i(\lambda)\rangle$ match that of the corresponding phc FQH state
up to $\xi_{\textrm{max}}\approx10$ at $\lambda=0.1$, but completely deviate at $\lambda=0.5$.
Our results clearly show that the adiabatic continuity indeed does not hold for fermions
at $\nu=\frac{4}{5}$.

We also find a similar phase transition for fermions at $\nu=\frac{2}{3}$.
However, $\nu=\frac{2}{3}$ is probably on the border between competing compressible states and FCI states
(see Ref. \onlinecite{andreas} for such study in checkerboard lattice), thus it is quite likely
that the appearance of this phase transition may depend on the system size and
the shape of the samples. On the contrary, we expect the $\nu=\frac{4}{5}$ results showing a clear phase transition to be robust to such details.

\section{Discussion}\label{discussion}
In this paper, we have investigated the interpolation between the FCI states and
FQH states with the help of appropriately gauge-fixed Wannier wave functions.\cite{qi,gaugefixing} By demonstrating an almost constant gap and a large overlap during the interpolation, we provide the strong evidence that both Abelian and non-Abelian FCI states are adiabatically connected to their corresponding FQH (Laughlin and Moore-Read, respectively) states for fermions as well as for bosons.
The method used here may be seen as an improved version of the first study of adiabatic continuity,\cite{moller} which studied $\nu=\frac{1}{2}$ bosons, by utilizing the recent gauge-fixing insights of Ref. \onlinecite{gaugefixing}. It also provides a more direct and quantitative comparison between FCI and FQH than the like-wise elegant connection recently established via relating Chern bands to the Landau bands of the Hofstadter problem in Ref. \onlinecite{AdiabaticContinuity2}.

To underscore the non-triviality of our results, we have also considered fermions at filling factor $\nu=\frac{4}{5}$, for which there is no FCI state due to the particle-hole symmetry breaking\cite{andreas} in the Chern band (which is absent in a Landau level). The overlap and gap indeed drop drastically during the interpolation reflecting a phase transition. In Ref. \onlinecite{enhancing}, it was pointed out that  the particle-hole symmetry can be explicitly restored by adding a single-particle term to the standard normal-ordered Hamiltonian, $H_{\textrm{FCI}}$, which is used in most numerical studies including this work.\cite{disp} By examining the adiabatic continuity using the resulting particle-hole-symmetric FCI Hamiltonian, we find that the results are significantly less universal and that they crucially depend on details such as the system size, the tight-binding parameters, and the interaction we choose.

We have also given a report of the orbital entanglement spectrum (OES) of FCI states
based on the orbital cut in the localized Wannier basis. The low-lying parts of the OES
of FCI states match those of corresponding ideal model FQH states very well. By comparing the OES of FCI states with those of FQH states obtained from realistic interaction in Landau levels, we find that FCI states are closer to the ideal FQH states. The analysis of the OES generalizes earlier works on FQH states on the torus, and thereby provides an appealing picture of the FCI OES as composed of two CFT spectra with opposite chirality.\cite{Lauchli,Zhao} Invoking the bulk-edge correspondence,\cite{LiH,bulkedge1,bulkedge2,bulkedge3} our results also provide a glimpse of the gapless edge physics of the FCI phases.

Our work invites a number of interesting future directions. Perhaps most interestingly, it suggests a natural generalization studying FCI states in Chern bands with higher Chern number. Indeed, novel series of FCI states with arbitrary Chern number, $|\mathcal{C}|=N$, have recently been observed in numerics.\cite{ChernN,hcnonab} While these new states might correspond to appropriately symmetrized versions of multi-component FQH states,\cite{ChernN,hcnonab,cn} such ideas need to be substantiated by further investigations and more direct comparisons as would be possible within the framework used here.

Another important issue would be if the present formalism might have bearing for the is the development of a pseudopotential formalism for fractional Chern insulators. At present, there are two approaches,\cite{andreas,fcipseudo} one of which is built on the original (non-gauge-fixed) Wannier basis construction,\cite{fcipseudo} leading to apparently diverging predictions.

\acknowledgments
We acknowledge A. L\"auchli for several related collaborations. E.J.B. is supported by the Alexander von Humboldt foundation.
Z.L. is supported by the China Postdoctoral Science Foundation Grant No. 2012M520149.

\begin{figure}[h!]
\centerline{\includegraphics[height=7.5cm,width=0.8\linewidth]{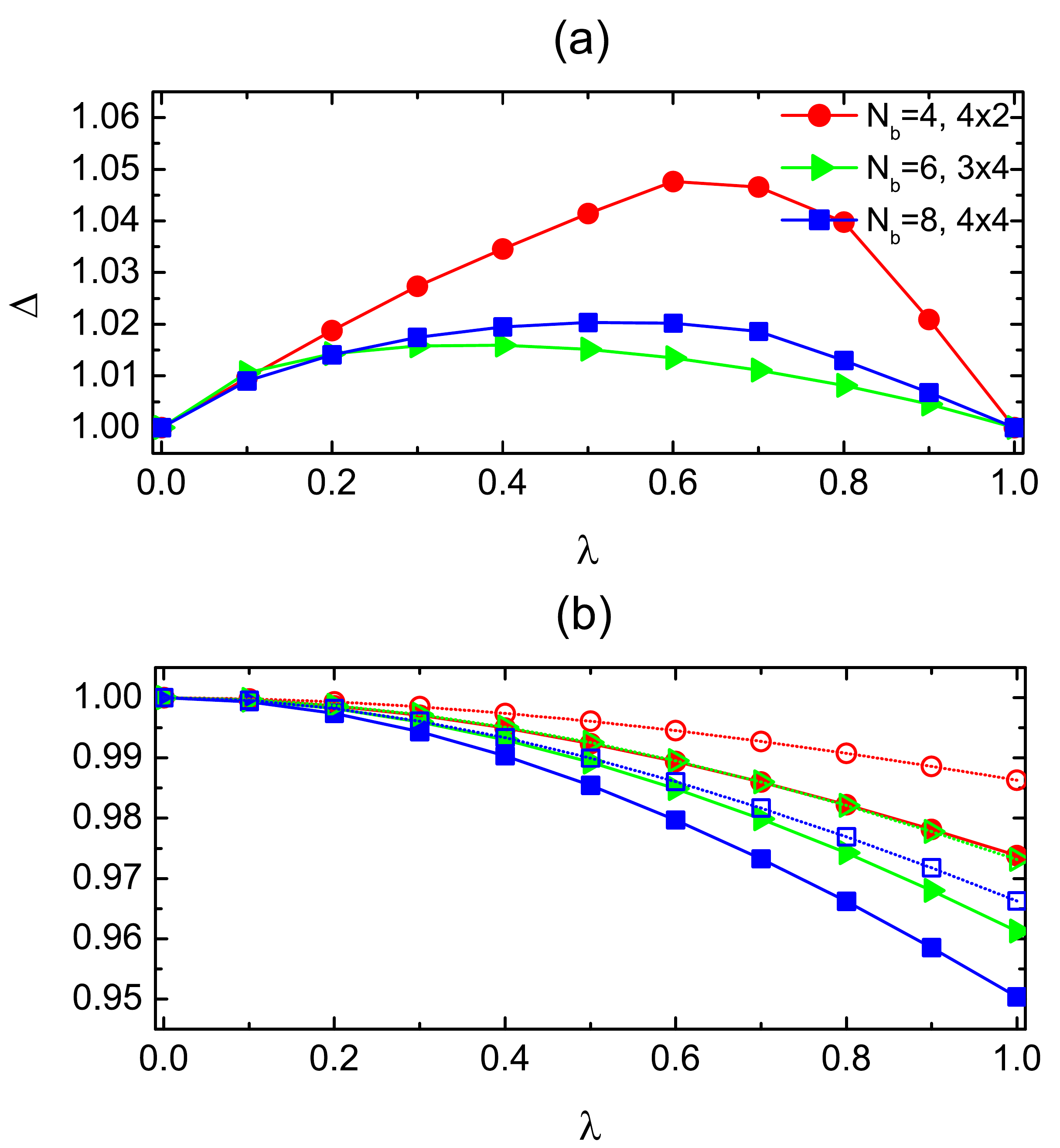}}
\caption{(Color online) Results of the interpolation Eq. (\ref{interpolate})
for bosons at $\nu=\frac{1}{2}$ for $N_b=4$ (red dot), $N_b=6$ (green triangle), and $N_b=8$ (blue square).
In the FCI part, the lattice size is $N_1\times N_2=4\times2$,
$N_1\times N_2=3\times4$ and $N_1\times N_2=4\times4$, respectively; and
$\lambda_1=1$.
(a) The energy gap $\Delta$ does not close for any intermediate $\lambda$.
(b) The total overlap $\mathcal{O}_{\textrm{tot}}$ (filled symbol, solid line) and
the average weight $\overline{\mathcal{W}}$ (empty symbol, dotted line) are still close to 1 at $\lambda=1$.
All of those demonstrate that the continuity holds for bosons at $\nu=\frac{1}{2}$.
}
\label{boslau}
\end{figure}

\begin{figure*}[t!]
\centerline{\includegraphics[height=10.5cm,width=0.7\linewidth]{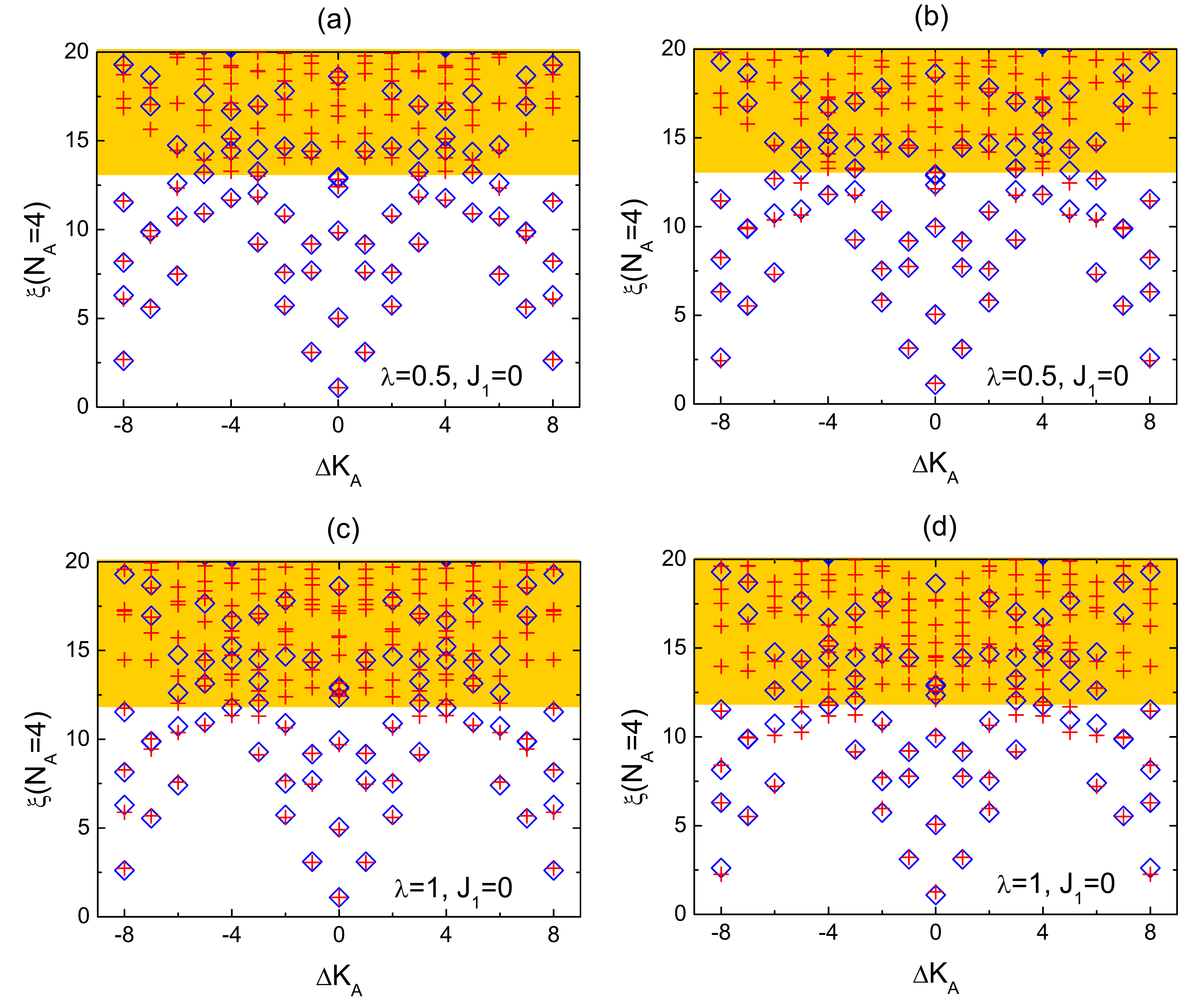}}
\caption{(Color online) The orbital entanglement spectra (OES) of exact bosonic Laughlin states (blue diamond) and
the projected nearly-degenerate states $|\Psi_{\textrm{prj}}^{J_1}(\lambda)\rangle$ (red cross)
at $\nu=\frac{1}{2}$, $N_b=8$. The lattice size is $N_1\times N_2=4\times4$ and $\lambda_1=1$ for the FCI part.
Both of the two nearly-degenerate states are in $J_1=0$ sector.
In the left column, we consider the one with lower energy and project it in
$K_1=0$ sector.
In the right column, we consider the one with higher energy and project it in
$K_1=8$ sector.
In (a) and (b), $\lambda=0.5$.
(a): The unprojected $|\Psi^{J_1}(\lambda)\rangle$ has weight $\mathcal{W}\approx0.79721$ on $K_1=0$ sector
and $\mathcal{W}\approx0.19096$ on $K_1=8$ sector. The overlap with the Laughlin state
$\mathcal{O}=\frac{1}{2}(|\langle\Psi_{\textrm{Lau}}^{K_1=0}|\Psi^{J_1=0}(\lambda)\rangle|^2+
|\langle\Psi_{\textrm{Lau}}^{K_1=8}|\Psi^{J_1=0}(\lambda)\rangle|^2)$ is 0.98264.
(b): The unprojected $|\Psi^{J_1}(\lambda)\rangle$ has weight $\mathcal{W}\approx0.19276$ on $K_1=0$ sector
and $\mathcal{W}\approx0.79902$ on $K_1=8$ sector. The overlap with the Laughlin state
$\mathcal{O}=\frac{1}{2}(|\langle\Psi_{\textrm{Lau}}^{K_1=0}|\Psi^{J_1=0}(\lambda)\rangle|^2+
|\langle\Psi_{\textrm{Lau}}^{K_1=8}|\Psi^{J_1=0}(\lambda)\rangle|^2)$ is 0.98825.
In (c) and (d), $\lambda=1$.
(c): The unprojected $|\Psi^{J_1}(\lambda)\rangle$ has weight $\mathcal{W}\approx0.75966$ on $K_1=0$ sector
and $\mathcal{W}\approx0.19968$ on $K_1=8$ sector. The overlap with the Laughlin state
$\mathcal{O}=\frac{1}{2}(|\langle\Psi_{\textrm{Lau}}^{K_1=0}|\Psi^{J_1=0}(\lambda)\rangle|^2+
|\langle\Psi_{\textrm{Lau}}^{K_1=8}|\Psi^{J_1=0}(\lambda)\rangle|^2)$ is 0.93967.
(d): The unprojected $|\Psi^{J_1}(\lambda)\rangle$ has weight $\mathcal{W}\approx0.20603$ on $K_1=0$ sector
and $\mathcal{W}\approx0.76720$ on $K_1=8$ sector. The overlap with the Laughlin state
$\mathcal{O}=\frac{1}{2}(|\langle\Psi_{\textrm{Lau}}^{K_1=0}|\Psi^{J_1=0}(\lambda)\rangle|^2+
|\langle\Psi_{\textrm{Lau}}^{K_1=8}|\Psi^{J_1=0}(\lambda)\rangle|^2)$ is 0.96111.
The orange shadows indicate the generic levels in the OES of $|\Psi_{\textrm{prj}}^{J_1}(\lambda)\rangle$
which deviate from the levels of the exact Laughlin state.
\label{boslauoes}}
\end{figure*}

\appendix \label{app}
\section{Bosons at $\nu=\frac{1}{2}$}
In this section we focus on the continuity problem of bosons at filling factor $\nu=\frac{1}{2}$.
On the FQH side, we choose the Hamiltonian as $H_{\textrm{FQH}}=\sum_{i<j}\delta^2(\textbf{r}_i-\textbf{r}_j)$.
Then the ground states are exact two-fold degenerate bosonic Laughlin states with zero energy. On the FCI side, the Hamiltonian is set as
the on-site interaction $H_{\textrm{FCI}}=\sum_{i }n_i(n_i-1)$.
The ground states are two nearly-degenerate states separated by a gap
from the excited states.
We set $\alpha=\pi/3$ also for the twisted torus.

We find that there are two nearly-degenerate states $|\Psi^i(\lambda)\rangle$ for each
$\lambda\in[0,1]$. The evolution of the energy gap [Fig. \ref{boslau}(a)], total overlap and average weight
[Fig. \ref{boslau}(b)] is similar with that for the $\nu=\frac{1}{3}$ fermionic Laughlin phase.
However, both of the total overlap and average weight are smaller ($\mathcal{O}_{\textrm{tot}}\approx0.950$
and $\overline{\mathcal{W}}\approx0.9663$ at $\lambda=1$ for our largest system size $N_b=8$).

We consider the OES for $N_b=8$ (the lattice size is $N_1\times N_2=4\times4$
for the FCI part). In the continuum, the thin-torus configuration of the two bosonic Laughlin states are
(for $N_b=8$, $N_s=16$)
\begin{eqnarray}
&0101|\textbf{01010101}|0101,&\nonumber\\
&1010|\textbf{10101010}|1010&\nonumber.
\label{boslautt}
\end{eqnarray}
Their total momentum is $K_1=0$ and $K_1=8$, respectively.
The two $|\Psi^i(\lambda)\rangle$ are both in $J_1=0$ sector, so
they mix with each other and do not have a good one-to-one correspondence with the two Laughlin states.
This means that each $|\Psi^i(\lambda)\rangle$ has weight on $K_1=0$ and $K_1=8$ sectors simultaneously.
However, we can still project one $|\Psi^i(\lambda)\rangle$
in $K_1=0$ sector and project the other in $K_1=8$ sector.
In Fig. \ref{boslauoes}, we can see that the low-lying part of the OES of $|\Psi^i_{\textrm{prj}}(\lambda)\rangle$
also match that of the exact Laughlin state very well. Of course $\xi_{\textrm{max}}$ here
($\xi_{\textrm{max}}\approx13.2$ at $\lambda=0.5$ and $\xi_{\textrm{max}}\approx11.2$ at $\lambda=1$)
is lower than that in the $\nu=\frac{1}{3}$ fermionic Laughlin case due to the lower overlap.
However, all of our results strongly support that the FQH states are adiabatically connected to FCI states
for bosons at $\nu=\frac{1}{2}$.

\begin{figure}
\centerline{\includegraphics[height=7.5cm,width=0.8\linewidth]{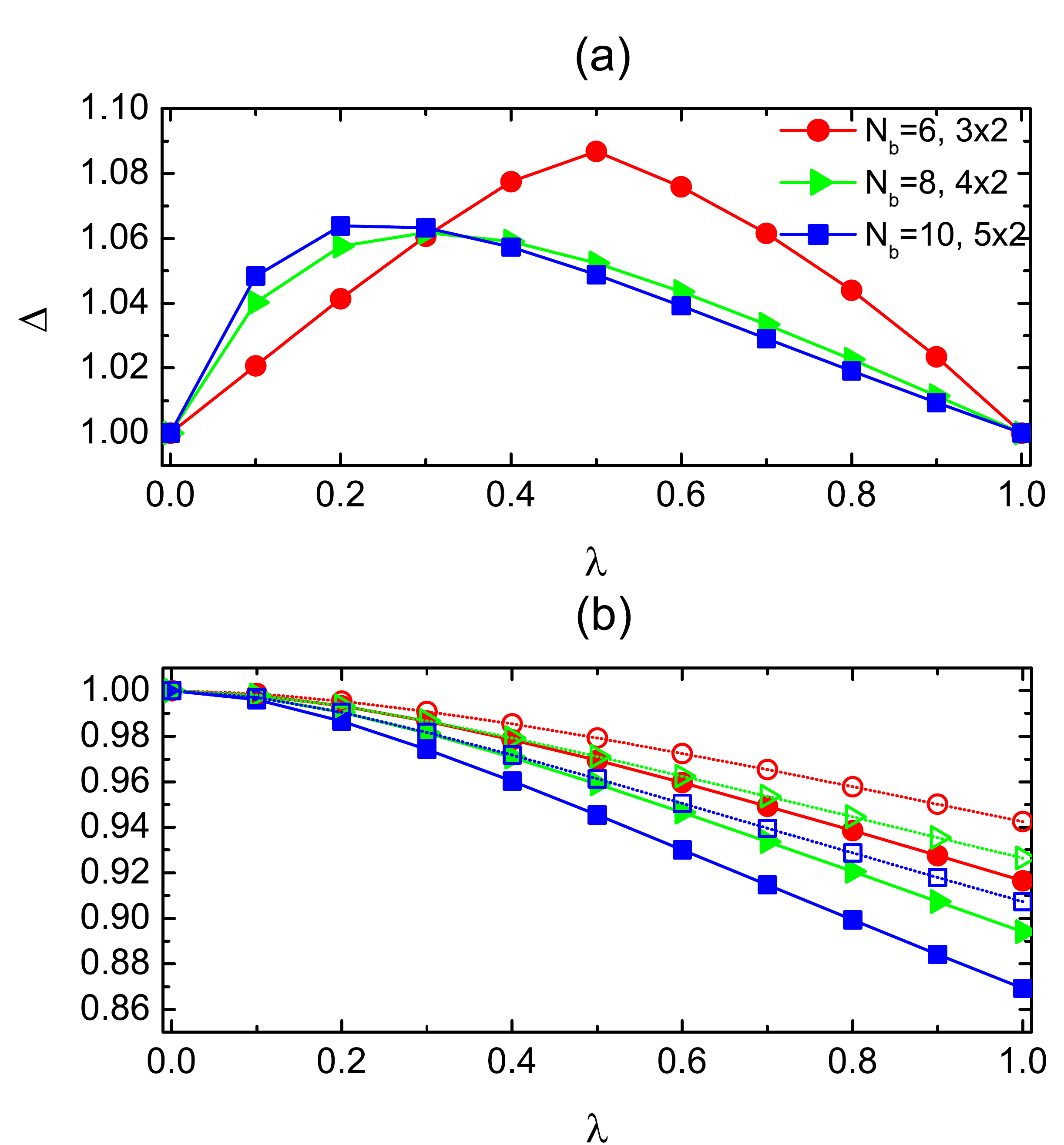}}
\caption{(Color online) Results of the interpolation Eq. (\ref{interpolate})
for bosons at $\nu=1$ for $N_b=6$ (red dot), $N_b=8$ (green triangle),
and $N_b=10$ (blue square). In the FCI part, the lattice size is $N_1\times N_2=3\times2$,
$N_1\times N_2=4\times2$ and $N_1\times N_2=5\times2$, respectively; and
$\lambda_1=0.8$.
(a) The energy gap $\Delta$ does not close for any intermediate $\lambda$.
(b) The total overlap $\mathcal{O}_{\textrm{tot}}$ (filled symbol, solid line) and
the average weight $\overline{\mathcal{W}}$ (empty symbol, dotted line) are still large at $\lambda=1$.
All of those demonstrate that the continuity holds for bosons at $\nu=1$.}
\label{bospf}
\end{figure}

\begin{figure}
\centerline{\includegraphics[height=9.5cm,width=0.7\linewidth]{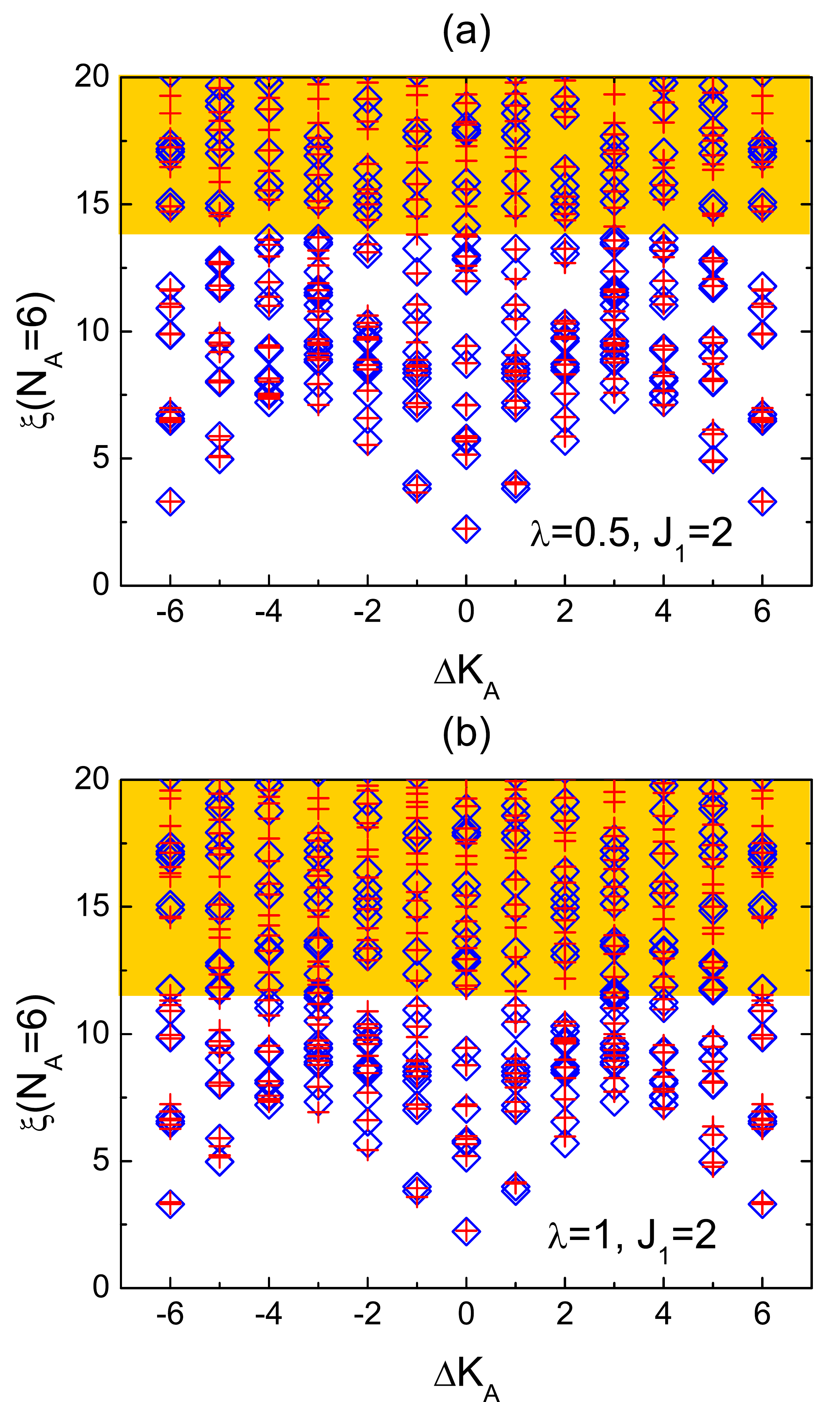}}
\caption{(Color online) (Color online) The orbital entanglement spectra (OES) of exact bosonic Moore-Read states (blue diamond) and
the projected nearly-degenerate state $|\Psi_{\textrm{prj}}^{J_1}(\lambda)\rangle$ (red cross)
at $\nu=1$, $N_b=12$.
$|\Psi_{\textrm{prj}}^{J_1}(\lambda)\rangle$ is in $J_1=2$ sector and corresponds to the Moore-Read state in
$K_1=6$ sector.
The lattice size is $N_1\times N_2=3\times4$ and $\lambda_1=0.8$ for the FCI part.
(a): $\lambda=0.5$. The unprojected $|\Psi^{J_1}(\lambda)\rangle$ has weight $\mathcal{W}\approx0.95860$ on $K_1=6$ sector.
The overlap with the Moore-Read state $\mathcal{O}=|\langle\Psi_{\textrm{MR}}^{K_1=6}|\Psi^{J_1=2}(\lambda)\rangle|^2$ is 0.93707.
(b): $\lambda=1$. The unprojected $|\Psi^{J_1}(\lambda)\rangle$ has weight $\mathcal{W}\approx0.89528$ on $K_1=6$ sector.
The overlap with the Moore-Read state $\mathcal{O}=|\langle\Psi_{\textrm{MR}}^{K_1=6}|\Psi^{J_1=2}(\lambda)\rangle|^2$ is 0.83976.
The orange shadows indicate the generic levels in the OES of $|\Psi_{\textrm{prj}}^{J_1}(\lambda)\rangle$
which deviate from the levels of the exact Moore-Read state.}
\label{bospfoes}
\end{figure}

\section{Bosons at $\nu=1$}

In this section we focus on the continuity problem of bosons at filling factor $\nu=1$.
On the FQH side, we choose the Hamiltonian as $H_{\textrm{FQH}}=\sum_{i<j<k}\delta^2(\textbf{r}_i-\textbf{r}_j)\delta^2(\textbf{r}_j-\textbf{r}_k)$.
Then the ground states are exact three-fold degenerate bosonic Laughlin states with zero energy. On the FCI side, the Hamiltonian is tactically chosen as
the on-site three-body interaction $H_{\textrm{FCI}}=\sum_{i }n_i(n_i-1)(n_i-2)$.
The ground states are three nearly-degenerate states separated by a gap
from the excited states.
We set $\alpha=2\pi/3$ for the twisted torus as was done for the fermion case at $\nu=\frac{1}{2}$.

We find that there are three nearly-degenerate states $|\Psi^i(\lambda)\rangle$ for each
$\lambda\in[0,1]$. The evolution of the energy gap [Fig. \ref{bospf}(a)], total overlap and average weight
[Fig. \ref{bospf}(b)] is similar with that for the $\nu=\frac{1}{2}$ fermionic Moore-Read phase.
However, both of the total overlap and average weight are smaller ($\mathcal{O}_{\textrm{tot}}\approx0.869$
and $\overline{\mathcal{W}}\approx0.9073$ at $\lambda=1$ for our largest system size $N_b=10$).

We consider the OES for $N_b=12$ (the lattice size is $N_1\times N_2=3\times4$
for the FCI part). In the continuum, the thin-torus configuration of the two bosonic Laughlin states are
(for $N_b=12$, $N_s=12$)
\begin{eqnarray}
&111|\textbf{111111}|111,&\nonumber\\
&020|\textbf{202020}|202\pm202|\textbf{020202}|020&\nonumber.
\label{boslautt}
\end{eqnarray}
Their total momentum is $K_1=6$ and $K_1=0$ (two-fold).
The three $|\Psi^i(\lambda)\rangle$ are in $J_1=0$ sector (two-fold) and
$J_1=2$ sector. Because the two states in $J_1=0$ sector will mix with each other,
we only concentrate on the single state in $J_1=2$ sector, which corresponds to
the $K_1=6$ Moore-Read state. Compared with the fermionic Moore-Read case, the asymmetry problem in the OES
is more serious (Fig. \ref{bospfoes}). Although the OES of $|\Psi^{J_1=2}(\lambda)\rangle$ does not precisely match that of exact bosonic
Moore-Read state, we can still find a relatively good correspondence between their OES levels up to $\xi_{\textrm{max}}\approx13.8$ at $\lambda=0.5$ and $\xi_{\textrm{max}}\approx11.5$ at $\lambda=1$.

\section{Further details on the Wannier basis construction}
\label{gaugedetails}
To make this paper self-contained, we give the prescription of how to fix the phases $\Phi(k_2)$ in the Wannier function $|X,k_2\rangle$ in this appendix. The results were first found in Ref. \onlinecite{gaugefixing}.
The essential point in fixing the phase is to make the connection $\langle X,k_2|\hat{Y}|X',k_2'\rangle$ between adjacent Wannier states
independent of $X$ and $k_2$. Here, $\hat{Y}=\sum_{k_1,k_2}|k_1,k_2\rangle A_2(k_1,k_2)\langle k_1,k_2+1|$ is the unitary
projected position operator in the $\textbf{e}_y$ direction, and adjacent Wannier states are defined by $j^{X',k_2'}=j^{X,k_2}+\mathcal{C}$.
Because $j^{X,k_2}=N_2X+\mathcal{C}k_2+\delta_2$, $k_2'=k_2+1$ (mod $N_2$) for two adjacent Wannier states.
If increasing from $k_2$ to $k_2+1$ does not cross the boundary of pBZ, $X'=X$. Otherwise, $X'=X+\mathcal{C}$.

From the definition of the Wannier state, one can obtain that
\begin{eqnarray}
\langle X,k_2|\hat{Y}|X',k_2'\rangle=e^{\textrm{i}[\Phi(k_2')-\Phi(k_2)]}A_2(0,k_2)\mathcal{U}_2(k_2),\nonumber\\
\label{c1}
\end{eqnarray}
where
\begin{eqnarray}
\mathcal{U}_2(k_2)&=&\frac{1}{N_1}\sum_{k_1=0}^{N_1-1}\frac{W(k_1,k_2)}{W'(k_1,k_2)},\nonumber
\end{eqnarray}
with
\begin{eqnarray}
W(k_1,k_2)&=&\frac{\prod_{\kappa=0}^{k_1-1}A_1(\kappa,k_2)}{\prod_{\kappa=0}^{k_1-1}A_1(\kappa,k_2+1)}
\frac{A_2(k_1,k_2)}{A_2(0,k_2)},\nonumber
\end{eqnarray}
$W'(k_1,k_2)=[\mu_1(k_2)]^{k_1}$, and
\begin{eqnarray}
\mu_1(k_2)&=&\left\{
\begin{array}{cccc}
\frac{\lambda_1(k_2)}{\lambda_1(k_2+1)}, k_2+1\in \textrm{pBZ}\\
e^{2\pi \textrm{i}\mathcal{C}/N_1}\frac{\lambda_1(k_2)}{\lambda_1(k_2+1)}, \textrm{otherwise}
\end{array}
\right..\nonumber
\end{eqnarray}
The product of $\langle X,k_2|\hat{Y}|X',k_2'\rangle$ has a very simple form,
\begin{eqnarray}
\prod_{X=0}^{N_1-1}\prod_{k_2=0}^{N_2-1}\langle X,k_2|\hat{Y}|X',k_2'\rangle
=\Bigg[W_2(0)\prod_{k_2=0}^{N_2-1}\mathcal{U}_2(k_2)\Bigg]^{N_1}.\nonumber\\
\label{c2}
\end{eqnarray}
Defining $U_2(k_2)=\mathcal{U}_2(k_2)/|\mathcal{U}_2(k_2)|$ and introducing a phase
$(\omega_2)^{N_2}=\prod_{k_2=0}^{N_2-1}U_2(k_2)$ with the argument angle in $(-\pi/N_2,\pi/N_2)$,
we can choose $\langle X,k_2|\hat{Y}|X',k_2'\rangle=\lambda_2(0)\omega_2|\mathcal{U}_2(k_2)|$,
which satisfies Eq. (\ref{c2}).
Finally, by comparing this choice with Eq. (\ref{c1}), we have
\begin{eqnarray}
e^{\textrm{i}[\Phi(k_2')-\Phi(k_2)]}=\frac{\lambda_2(0)}{A_2(0,k_2)}\frac{\omega_2}{U_2(k_2)}.
\label{c3}
\end{eqnarray}
We can choose $e^{\textrm{i}\Phi(0)}=1$ and recursively fix all phases $e^{\textrm{i}\Phi(k_2)}$
according to Eq. (\ref{c3}).


\begin{thebibliography}{99}
\bibitem{q_1=1}In particular, if $q_1=1$ in Eq. (\ref{psisr}), Eq. (\ref{wanniermap}) defines a direct correspondence between
$|\Psi_{\textrm{FQH}}(K_1,K_2)\rangle$ and $|\Psi_{\textrm{FQH}}^{\textrm{lat}}(K_1,K_2)\rangle$.
In this paper, we always set $q_1=1$ by choosing appropriate $N_1$ and $N_2$.
\end{thebibliography}
\end{document}